\documentclass[twocolumn,amsmath,amssymb,aps, physrev, superscriptaddress
]{revtex4-2}

\usepackage{graphicx}
\usepackage{dcolumn}
\usepackage{bm}
\usepackage{hyperref}
\usepackage[mathlines]{lineno}
\usepackage{amsmath}
\usepackage{siunitx}
\usepackage{physics}
\usepackage{microtype}
\usepackage{tabularx}
\usepackage[capitalise]{cleveref}

\begin{document}

\preprint{APS/123-QED}

\title{\textbf{Exploring $\alpha$- and $\beta$-decay-induced quenching of the $^{229}$Th nuclear-clock isomer in solid-state hosts} 
}%

\author{Y.~Elskens}
\email{yens.elskens@kuleuven.be}
\affiliation{KU Leuven, Instituut voor Kern- en Stralingsfysica, 3001 Leuven, Belgium}

\author{M.~Athanasakis-Kaklamanakis}
\affiliation{KU Leuven, Instituut voor Kern- en Stralingsfysica, 3001 Leuven, Belgium}
\author{S.~Arasada Pradeep}
\affiliation{Institute of Physics -- Johannes Gutenberg University Mainz, 55099 Mainz, Germany}
\author{M.~Au}
\affiliation{CERN SY-STI, Geneva, 1211, Switzerland}
\author{S.~Bara}
\affiliation{KU Leuven, Instituut voor Kern- en Stralingsfysica, 3001 Leuven, Belgium}
\affiliation{Université de Caen Normandie, ENSICAEN, CNRS/IN2P3, LPC CAEN, F-14000 Caen, France}
\affiliation{Grand Accélérateur National d'Ions Lourds (GANIL), CEA/DRF-CNRS/IN2P3, F-14076 Caen, France}
\author{M.~Bartokos}
\affiliation{Vienna Center for Quantum Science and Technology, Atominstitut, TU Wien, 1020 Vienna, Austria}
\author{K.~Beeks}
\affiliation{Vienna Center for Quantum Science and Technology, Atominstitut, TU Wien, 1020 Vienna, Austria}
\author{C.~Bernerd}
\affiliation{KU Leuven, Instituut voor Kern- en Stralingsfysica, 3001 Leuven, Belgium}
\author{B.~Biesmans}
\affiliation{KU Leuven, Quantum Solid State Physics, 3001 Leuven, Belgium}
\author{S.~Casci}
\affiliation{KU Leuven, Instituut voor Kern- en Stralingsfysica, 3001 Leuven, Belgium}
\author{P.~Chhetri}%
\affiliation{Department of Chemistry -- TRIGA Site, Johannes Gutenberg-Universität Mainz, 55099 Mainz, Germany}
\author{K.~Chrysalidis}
\affiliation{CERN SY-STI, Geneva, 1211, Switzerland}
\author{A.~Claessens}
\affiliation{KU Leuven, Instituut voor Kern- en Stralingsfysica, 3001 Leuven, Belgium}
\author{T.~E.~Cocolios}
\affiliation{KU Leuven, Instituut voor Kern- en Stralingsfysica, 3001 Leuven, Belgium}
\author{J.~G.~Correia}
\affiliation{Centro de Ciências e Tecnologias Nucleares, Departamento de Engenharia e Ciências Nucleares, Instituto Superior Técnico, Universidade de Lisboa, 2695-066 Bobadela LRS, Portugal}

\author{A.~R.~G.~Costa}
\affiliation{Centro de Ciências e Tecnologias Nucleares, Departamento de Engenharia e Ciências Nucleares, Instituto Superior Técnico, Universidade de Lisboa, 2695-066 Bobadela LRS, Portugal}

\author{H.~De Witte}
\affiliation{KU Leuven, Instituut voor Kern- en Stralingsfysica, 3001 Leuven, Belgium}
\author{S.~B.~Diewald}
\affiliation{Department of Chemistry -- TRIGA Site, Johannes Gutenberg-Universität Mainz, 55099 Mainz, Germany}
\author{Ch.~E.~D\"ullmann}
\affiliation{Department of Chemistry -- TRIGA Site, Johannes Gutenberg-Universität Mainz, 55099 Mainz, Germany}
\affiliation{GSI Helmholtzzentrum für Schwerionenforschung GmbH, 64291 Darmstadt, Germany}
\affiliation{Helmholtz Institute Mainz, 55099 Mainz, Germany}

\author{R.~Ferrer}
\affiliation{KU Leuven, Instituut voor Kern- en Stralingsfysica, 3001 Leuven, Belgium}
\author{R.~Heinke}
\affiliation{CERN SY-STI, Geneva, 1211, Switzerland}
\author{G.~Holthoff}
\affiliation{Ludwig-Maximilians-Universit\"at M\"unchen, 85748 Garching, Germany}
\author{F.~Ivandikov}
\affiliation{KU Leuven, Instituut voor Kern- en Stralingsfysica, 3001 Leuven, Belgium}
\author{Yu.~Kudryavtsev}
\affiliation{KU Leuven, Instituut voor Kern- en Stralingsfysica, 3001 Leuven, Belgium}
\author{U.~K\"oster}
\affiliation{Institut Laue Langevin, 38042 Grenoble, France}
\author{S.~Kraemer}
\affiliation{KU Leuven, Instituut voor Kern- en Stralingsfysica, 3001 Leuven, Belgium}
\author{M.~Laatiaoui}
\affiliation{Department of Chemistry -- TRIGA Site, Johannes Gutenberg-Universität Mainz, 55099 Mainz, Germany}
\affiliation{Grand Accélérateur National d'Ions Lourds (GANIL), CEA/DRF-CNRS/IN2P3, F-14076 Caen, France}
\author{R.~Lica}
\affiliation{Horia Hulubei National Institute for R\&D in Physics and Nuclear Engineering, RO-077125 Bucharest, Romania}
\author{C.~Merckling}
\affiliation{IMEC, 3001 Leuven, Belgium}
\author{J.~Moens}
\affiliation{KU Leuven, Quantum Solid State Physics, 3001 Leuven, Belgium}
\author{I.~Morawetz}
\affiliation{Vienna Center for Quantum Science and Technology, Atominstitut, TU Wien, 1020 Vienna, Austria}
\author{D.~Moritz}
\affiliation{Ludwig-Maximilians-Universit\"at M\"unchen, 85748 Garching, Germany}
\author{L.~M.~C.~Pereira}
\affiliation{KU Leuven, Quantum Solid State Physics, 3001 Leuven, Belgium}
\author{S.~V.~Pineda}
\affiliation{KU Leuven, Instituut voor Kern- en Stralingsfysica, 3001 Leuven, Belgium}
\affiliation{Université de Caen Normandie, ENSICAEN, CNRS/IN2P3, LPC CAEN, F-14000 Caen, France}
\affiliation{Grand Accélérateur National d'Ions Lourds (GANIL), CEA/DRF-CNRS/IN2P3, F-14076 Caen, France}
\author{S.~Raeder}
\affiliation{GSI Helmholtzzentrum für Schwerionenforschung GmbH, 64291 Darmstadt, Germany}
\author{S.~Rothe}
\affiliation{CERN SY-STI, Geneva, 1211, Switzerland}
\author{S.~Sabrieva}
\affiliation{Vienna Center for Quantum Science and Technology, Atominstitut, TU Wien, 1020 Vienna, Austria}
\author{M.~Satrazani}
\affiliation{KU Leuven, Instituut voor Kern- en Stralingsfysica, 3001 Leuven, Belgium}
\author{F.~Schaden}
\affiliation{Vienna Center for Quantum Science and Technology, Atominstitut, TU Wien, 1020 Vienna, Austria}
\author{K.~Scharl}
\affiliation{Ludwig-Maximilians-Universit\"at M\"unchen, 85748 Garching, Germany}
\author{T.~Schumm}
\affiliation{Vienna Center for Quantum Science and Technology, Atominstitut, TU Wien, 1020 Vienna, Austria}
\author{S.~Stegemann}
\affiliation{CERN SY-STI, Geneva, 1211, Switzerland}
\author{J.~Stricker}
\affiliation{Department of Chemistry -- TRIGA Site, Johannes Gutenberg-Universität Mainz, 55099 Mainz, Germany}
\affiliation{Helmholtz Institute Mainz, 55099 Mainz, Germany}
\author{T.~Teschler}
\affiliation{Ludwig-Maximilians-Universit\"at M\"unchen, 85748 Garching, Germany}
\author{P.~G.~Thirolf}
\affiliation{Ludwig-Maximilians-Universit\"at M\"unchen, 85748 Garching, Germany}
\author{P.~Van den Bergh}
\affiliation{KU Leuven, Instituut voor Kern- en Stralingsfysica, 3001 Leuven, Belgium}
\author{P.~Van Duppen}
\affiliation{KU Leuven, Instituut voor Kern- en Stralingsfysica, 3001 Leuven, Belgium}
\author{A.~Vantomme}
\affiliation{KU Leuven, Quantum Solid State Physics, 3001 Leuven, Belgium}
\author{R.~Villarreal}
\affiliation{KU Leuven, Quantum Solid State Physics, 3001 Leuven, Belgium}
\author{L. von der Wense}
\affiliation{Institute of Physics -- Johannes Gutenberg University Mainz, 55099 Mainz, Germany}
\author{U.~Wahl}
\affiliation{Centro de Ciências e Tecnologias Nucleares, Departamento de Engenharia e Ciências Nucleares, Instituto Superior Técnico, Universidade de Lisboa, 2695-066 Bobadela LRS, Portugal}
\author{Y.~Wang}
\affiliation{Institute of Physics -- Johannes Gutenberg University Mainz, 55099 Mainz, Germany}
\author{M.~Wiesinger}
\affiliation{Ludwig-Maximilians-Universit\"at M\"unchen, 85748 Garching, Germany}
\author{Z.~Yue}
\affiliation{CERN SY-STI, Geneva, 1211, Switzerland}
\author{F.~Zacherl}
\affiliation{Institute of Physics -- Johannes Gutenberg University Mainz, 55099 Mainz, Germany}



\date{\today}

\begin{abstract}
The radiative decay dynamics of an ensemble of $^{229\mathrm{m}}$Th nuclei embedded in CaF$_2$ and MgF$_2$ is investigated. The isomer is populated through $\beta$ decay of $^{229}$Ac following ion implantation, and its radiative decay is detected using vacuum-ultraviolet spectroscopy and measured as a function of time. This allows to identify and quantify the quenching of the radiative-decay signal induced by $\alpha$ or $\beta$ radiation. The quenching probability density is determined in different CaF$_2$ crystals and in a MgF$_2$ crystal, revealing differences up to two orders of magnitude between the investigated samples and a strong dependence on the host material and defect densities. The results support a microscopic mechanism mediated by charge carriers in which electronic excitations created by the decay radiation are captured near Th defects, thereby favoring non-radiative decay channels.
\end{abstract}

\maketitle


\section{Introduction}
The isomeric state in the $^{229}$Th nucleus has been the subject of extensive research over the last few decades. With its unique proximity to the ground state (at 8.356~eV), this isomeric state can be directly populated by excitation with vacuum-ultraviolet (VUV) laser light, making this nuclear transition an ideal frequency-reference transition for developing a nuclear clock~\cite{peik_nuclear_2003}. As nuclear transitions are better shielded from environmental factors than electronic transitions, nuclear clocks are expected to improve on accuracy compared to atomic clocks. Moreover, the possibility to use a solid-state host allows to probe thorium densities on the order of $10^{18}\,\si{\per\cm\cubed}$~\cite{tiedau_laser_2024}, representing an increase of about 11 orders of magnitude compared to ion traps, reducing the quantum projection noise substantially. Additionally, nuclear clocks are particularly coveted as a quantum sensor for ultralight dark-matter, and their predicted sensitivity to variations of fundamental constants is expected to be exceptionally high compared to state-of-the-art optical atomic clocks~\cite{peik_nuclear_2021, thirolf_thorium_2024,caputo_2025_sensitivity,arakawa_arXiv_2026,toscani_2026_nuclear_clock_arxiv} 

After the first direct experimental evidence for the existence of a low-lying isomer in $^{229}$Th~\cite{vonderwense_detection_2016} and following the approach described in~\cite{verlinde_alternative_2019}, the radiative decay of $^{229\mathrm{m}}$Th was observed with a vacuum-ultraviolet spectrometer by populating the isomer through the $\beta$ decay of $^{229}$Ac~\cite{kraemer_observation_2023}. This was achieved by implanting radioactive $^{229}$Ra ions, which subsequently $\beta$-decay to $^{229}$Ac, into CaF$_2$ and MgF$_2$ crystals. These measurements significantly reduced the uncertainty on the transition energy to the $10^{-3}$ regime and helped identify other suitable solid-state host materials, such as LiSrAlF$_6$~\cite{pineda_radiative_2025}. This paved the way to study the isomer through laser excitation from the ground state in CaF$_2$~\cite{tiedau_laser_2024} and LiSrAlF$_6$~\cite{elwell_laser_2024}. Subsequently, the laser excitation in CaF$_2$ with a VUV frequency comb was achieved, further reducing the relative uncertainty on the transition energy to the $10^{-12}$ regime, and allowing to resolve the nuclear hyperfine structure~\cite{zhang_frequency_2024}. 

A central challenge in the development of a nuclear clock is the suppression of the internal conversion (IC) decay channel. For neutral thorium, the first ionization potential lies at 6.306879(14) eV~\cite{claessens2025_ionTh_PRA}, which is below the isomer's excitation energy. As a result, the IC decay channel is open in atomic $^{229\mathrm{m}}$Th and a half-life of $7\pm 1\,\si{\micro\second}$ in a metallic environment has been reported~\cite{seiferle2017_lifetime_PRL}. Using $^{229}$Th dopants embedded in wide-band-gap crystals allows this otherwise dominant decay channel to be suppressed, provided that the local electronic spectrum at the Th site does not provide an efficient energy-conserving non-radiative pathway at the nuclear transition energy $E_{\mathrm{iso}}$. In an idealized single-particle picture, this condition is often expressed as a local gap between the highest occupied and lowest unoccupied electronic states that exceeds $E_{\mathrm{iso}}$, and is sometimes referred to as ``preservation of the band gap''. In a real crystal, however, the relevant condition is more microscopic: localized defect or impurity states, charge-compensation structures, and carrier-trapping configurations can modify both the local electronic spectrum and the coupling matrix elements at the Th site. Such states quench the isomer efficiently only if they provide occupied-to-empty electronic spectral weight near $E_{\mathrm{iso}}$ with sufficient coupling to the nuclear transition. Sub-gap states that are strongly detuned from $E_{\mathrm{iso}}$, weakly coupled to the Th nucleus, or unfavorably occupied are therefore expected to contribute only weakly. The overall ability to suppress IC is therefore strongly dependent on the crystal material and on the local structural-electronic configuration of the Th impurity~\cite{pineda_radiative_2025}.

The relevance of the crystalline environment on the decay of an ensemble of $^{229\mathrm{m}}$Th nuclei became apparent when a reduction of the VUV radiative-decay signal was observed within CaF$_2$~\cite{hiraki_Xray_2024,guan_xray_2026,schaden_LIQ_2025} and LiSrAlF$_6$~\cite{terhune2025photo}. This so-called quenching was found to be induced by X-rays and laser photons, and  the intensity of such an effect was found to be dependent on flux, wavelength and temperature. Additionally, a dependency of the quenching on the doping site was observed by performing laser M\"ossbauer spectroscopy on the isomer embedded in CaF$_2$~\cite{hiraki2025lasermossbauer}. The origin of this mechanism has been attributed to processes that modify the local charge state of a Th-related defect, for example by capturing a mobile electron in a Th-coupled local state~\cite{terhune2025photo,guan_xray_2026}. Once occupied, such a state or local excitation can act as the initial state for IC into empty host or defect states, to the extent that the corresponding local electronic transition has spectral weight at $E_{\mathrm{iso}}$ and sufficient Th-projected character entering the nuclear-electronic coupling matrix element. Beyond its fundamental interest, this mechanism could be of practical importance, as it has been identified as a means to efficiently relax the excited-state population to the ground state~\cite{hiraki_Xray_2024,schaden_LIQ_2025,terhune2025photo, hiraki2025lasermossbauer,guan_xray_2026}.

In addition to externally induced quenching, the radioactivity of $^{229}$Th itself represents an intrinsic source of quenching: the $\alpha$ and $\beta$ decays occurring in the host crystal can deposit energy locally and potentially trigger the quenching mechanism. This raises the possibility of self-quenching in realistic solid-state samples, particularly at high thorium densities. In this manuscript we report the first systematic study of quenching induced by $\alpha$ and $\beta$ radiation using the methodology reported in Refs.~\cite{kraemer_observation_2023,kraemer_setup_2023,pineda_radiative_2025} to observe the time behavior of the radiative decay of the isomer embedded in CaF$_2$ and MgF$_2$. The quenching effect will be quantified and evidence for at least two radiatively active microscopic configurations with different activity-induced quenching probabilities will be presented.

\section{Experimental technique}
\subsection{Radioactive ion beam production}
A radioactive $A=229$ ion beam was produced at the Isotope Separator On-Line DEvice (ISOLDE) facility at the European organisation for nuclear research (CERN)~\cite{Catherall_ISOLDE_2017}. At this facility, a ThC$_x$ target was heated to 2000 \textdegree C and irradiated with a 2\,\si{\micro\ampere} current of 1.4\,\si{\giga\electronvolt} protons delivered by the PS-Booster.  The created reaction products are stopped in the target, then diffuse towards the target surface and effuse into a hot cavity where they are surface ionized to the 1+ charge state, a process governed by the physico-chemical properties of the elements. These ionized products are accelerated to 30~\si{\kilo\electronvolt} and separated according to their mass-to-charge ratio by the General-Purpose mass Separator (GPS). For our purposes, an $A=229$ beam was selected, which is composed of the $^{229}$Fr$^+ \rightarrow ^{229}$Ra$^+ \rightarrow ^{229}$Ac$^+$ $\beta$-decay chain. It is dominated by $^{229}$Ra, whose implantation rates are typically two orders of magnitude higher than for $^{229}$Fr and $^{229}$Ac. 

The $A=229$ beam was implanted into a set of large-bandgap crystals, positioned on a sample-holder wheel in the VUV-spectroscopy setup~\cite{kraemer_observation_2023,kraemer_setup_2023}. After implantation, the wheel was turned by 180\textdegree, placing the sample in front of a set of adjustable entrance slits, which, for this experiment, were kept at a width of 2\,\si{\milli\meter}. The spectrometer itself (Resonance Ltd., VM180) consists of a Czerny-Turner setup, which collects the VUV light, diffracts it into its respective wavelength components and focuses it onto a solar-blind photo-multiplier tube (PMT, Hamamatsu R8487). A stepper motor then pushes a crank connected to the grating axis to scan the VUV regime for a signal. As reported in Refs.~\cite{kraemer_thesis_2022,kraemer_setup_2023}, a plasma source was used in an environment with a controlled air-flow to calibrate the wavelength, using characteristic peaks of molecular oxygen and nitrogen in the VUV regime. Using these molecular lines, a wavelength calibration of the grating position was performed, and the ROI (115-195 nm) could be scanned.

The present experiment required the production of two additional radioactive ion beams at masses $A=230$ and $A=220$. The $A=230$ beam consists of pure $\beta$ emitters $^{230}$Ra ($t_{1/2}=93\pm2$ min) and its daughter $^{230}$Ac ($t_{1/2}=122\pm3$ s), which, once implanted into the considered crystals, induce Cherenkov light. This allowed to model the Cherenkov contribution which dominates the background in the $A=229$ spectra. 

The $A=220$ beam consisted of the short-lived $^{220}$Fr ($t_{1/2}=27.4\pm 0.3$~s, $E_\alpha = 6677\pm 4\,\si{\kilo\electronvolt}$~\cite{Morse2026NDS}), and was used to quantify the effect of $\alpha$ decay on the radiative-decay rate. Its decay chain reaches, through pure $\alpha$ decay, equilibrium with $^{212}$Bi ($t_{1/2}=60.55\pm0.06$~min, $Q_\beta = 2251.1\pm1.7\,\si{\kilo\electronvolt}$~\cite{Auranen2020NDS}, $E_\alpha = 6050.78\pm0.03\,\si{\kilo\electronvolt}$~\cite{Martin2007NDS}). From there, the decay chain splits up into a $\beta$ ($\rightarrow^{212}$Po, $t_{1/2}=45.1\pm0.6$~s, $E_\alpha = 11\,660\pm10\,\si{\kilo\electronvolt}$~\cite{Martin2007NDS}) and an $\alpha$ ($\rightarrow^{208}$Tl, $t_{1/2}=3.053\pm0.004$~min, $Q_\beta = 4998.9\pm1.8\,\si{\kilo\electronvolt}$~\cite{Martin2007NDS}) branch, whose daughters, respectively, $\alpha$ and $\beta$ decay to $^{208}$Pb.

To infer the amount of $^{229}$Th isomers implanted into each crystal, $\gamma$-ray spectroscopy was performed using a high purity germanium (HPGe) coaxial detector that was placed near the implantation point. Activity calibrated $\gamma$-ray sources of $^{60}$Co, $^{137}$Cs, $^{152}$Eu and $^{133}$Ba were used to determine the energy calibration and absolute efficiency of the HPGe detector. 
During the implantation, the intensity of the characteristic $\gamma$ rays of  $^{229}$Fr ($t_{1/2} = 50.2\pm2.0\,\si{\second}$, $E_\gamma = 1256\,\si{\kilo\electronvolt}$, $I_\gamma = 3.7\pm0.4\,\%$, $Q_\beta = 3106\pm16\,\si{\kilo\electronvolt}$~\cite{Kondev2026NDS}), $^{229}$Ra ($t_{1/2} = 4.0\pm0.2\,\si{\minute}$, $E_\gamma = 445\,\si{\kilo\electronvolt}$, $I_\gamma=10.7\pm 0.7\,\%$~\cite{CasciThesis229Th}, $Q_\beta = 1872\pm20\,\si{\kilo\electronvolt}$~\cite{Kondev2026NDS}) and $^{229}$Ac ($t_{1/2} = 62.7\pm0.5\,\si{\minute}$, $E_\gamma = 569\,\si{\kilo\electronvolt}$, $I_\gamma = 4.39\pm0.25\,\%$~\cite{CasciThesis229Th}, $Q_\beta = 1104\pm12\,\si{\kilo\electronvolt}$~\cite{Kondev2026NDS}) were monitored as a function of time. The beam intensity for all precursor isotopes was determined by fitting Bateman's equations to the time evolution of the $\gamma$-ray count rates, and correcting for $\gamma$-ray detection efficiency and absolute intensity. Note that the absolute $\gamma$-ray intensity of the transitions from $^{229}$Ra and $^{229}$Ac was deduced from an off-line $\alpha$-spectroscopy measurement of the implanted samples as reported in Refs.~\cite{pineda_radiative_2025,CasciThesis229Th}.
The $A=229$ radioactive beam was mainly composed of $^{229}$Ra and resulted in implantation rates for $^{229}$Ra varying between $3\cdot10^{7}$ and $1.5\cdot 10^{8}$ pps. Assuming typical implantation times ranging between $900\,\si{\second}$ and $3600\,\si{\second}$, a full width at half maximum of the implantation  depth profile of $\sim 7\,\si{\nano\meter}$, and a beam-spot size of $1.5\,\si{\milli\meter}\times2.7\,\si{\milli\meter}$ (as reported in~\cite{pineda_radiative_2025}), this delivers typical local $^{229}$Th densities of the order of $10^{19}\,\si{\per\centi\meter\cubed}$.

The implantation rates for $^{220}$Fr could not be determined in the same manner due to the low $\gamma$-ray intensities. Instead they were estimated from the current measured on the Faraday cup upstream of the setup. A calibration factor was obtained from the $A=229$ beam by comparing Faraday-cup readings to implantation rates determined via $\gamma$-ray spectroscopy, leading to an estimated implantation rate of $2\cdot 10^8$ pps for $^{220}$Fr.

\subsection{Crystal specifications}
A summary of the different samples used in the experiments and their specifications is provided in~\cref{tab:specs}. For CaF$_2$, two types of epitaxial thin-film crystals were tested as well as two types of bulk-material crystals. The two thin films were grown on a 0.75 mm thick Si(111) substrate at Imec (Leuven, Belgium), using molecular beam epitaxy (MBE). They were cut from the same wafer as the samples used in Ref.~\cite{pineda_radiative_2025}, and the labels `350' and `850' indicate the growth temperature in \textdegree C. 
\begin{table}[h!]
\nolinenumbers
\centering
\caption{Specifications on the used crystals. As the bulk-material crystals were circular in shape, their size is given by their diameter and thickness, while the size of the thin-film crystals is reported by length ($\ell$), width ($w$) and thickness ($t$).} \label{tab:specs}
\begin{ruledtabular}
    \begin{tabular}{cccc}
         Crystal & Size ($\ell\times w \times t$) & Substrate & Manufacturer \\
\hline
CaF$_2$ 350 & $2\,\mathrm{cm}\times 2\,\mathrm{cm} \times 50\,\mathrm{nm}$ & Si(111) & Imec   \\
CaF$_2$ 850 & $2\,\mathrm{cm}\times2\,\mathrm{cm}\times 50\,\mathrm{nm}$ & Si(111) & Imec  \\
CaF$_2$ standard & $\varnothing25.4\,\mathrm{mm} \times 5\,\mathrm{mm}$ & -- & Thorlabs \\
CaF$_2$ UV  & $\varnothing25.4\,\mathrm{mm} \times 2\,\mathrm{mm}$ & -- & Crystran \\
MgF$_2$ bulk & $\varnothing25.4\,\mathrm{mm} \times 5\,\mathrm{mm}$ & -- & Thorlabs \\
    \end{tabular}
\end{ruledtabular}
  \linenumbers
\end{table}

\begin{figure}
    \centering
    \includegraphics[width=\linewidth]{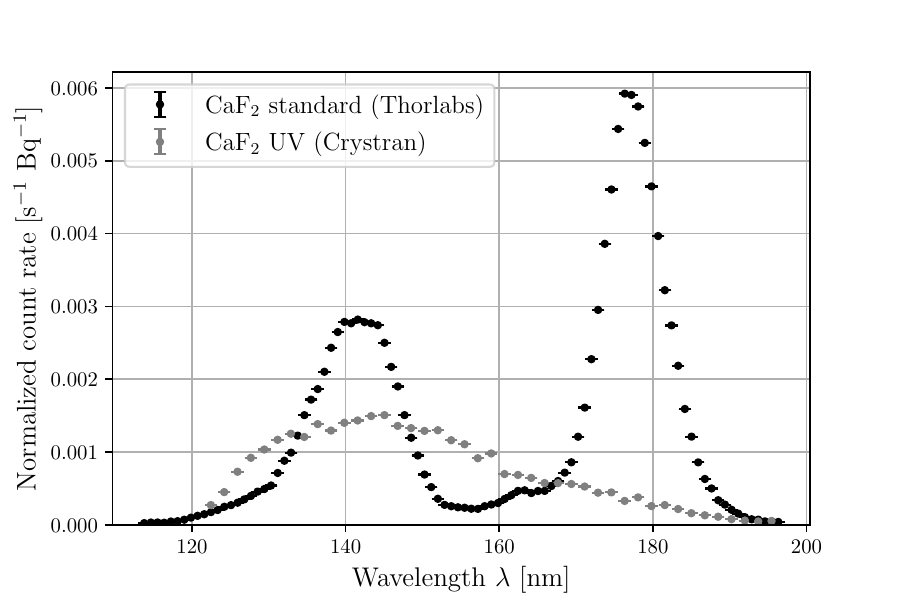}
    \caption{VUV spectra after $A=230$ implantations for the standard-grade CaF$_2$ and the UV-polished CaF$_2$ samples normalized by radioactivity and corrected for geometrical-efficiency differences. Broad radioluminescence peaks, which are present in the standard-grade CaF$_2$ at approximately 142 nm and 177 nm, are not observed in CaF$_2$ UV.}
    \label{fig:cherenkov_bulk}
\end{figure}

The CaF$_2$ UV-grade crystal from Crystran was, after manufacturing, polished to minimize UV absorption and scattering for better transmission in the VUV region. The VUV spectra obtained from both crystals following $A=230$ implantations (\cref{fig:cherenkov_bulk}), show  the presence of radioluminescence peaks in the CaF$_2$ standard grade sample from Thorlabs which are not observed in the CaF$_2$ UV-polished sample. It is believed that these peaks are caused by the activation of photon-emitting defect states, so-called color centers, by the radiation from the $A=230$ decay chain. While not linked to the UV-grade polishing technique, these $A=230$ spectra do indicate a lower defect density in CaF$_2$ UV compared to CaF$_2$ bulk.

\subsection{Radiative-decay spectra}
\begin{figure}
    \centering
    \includegraphics[width=\linewidth]{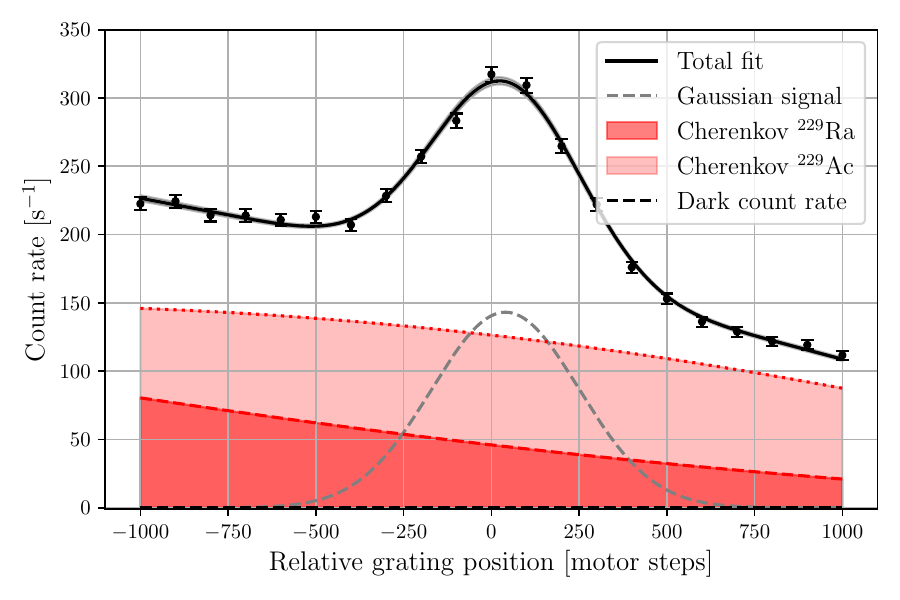}
    \caption{Example of a spectrum collected from a MgF$_2$ crystal, showing the polynomial and Gaussian contributions to the total spectrum, as described in \cref{eq: spectrum_description}. The centroid was reached $1100\,\si{\second}$ after the end of an $1800\,\si{\second}$ $A=229$ implantation.}
    \label{fig:spec_MgF2}
\end{figure}

To obtain information on the time behavior of the isomer's radiative decay, the samples were implanted with an $A=229$ beam for durations ranging between $900\,\si{\second}$ to $3600\,\si{\second}$, depending on the production yield. After implantation and positioning of the sample in front of the entrance slits, the region  of interest around the 148 nm wavelength peak was scanned repeatedly. This was done by moving the grating in wavelength steps of 0.87 nm, collecting VUV counts at each step for 10 s. The global shape of the spectra was described (as in Refs.~\cite{kraemer_thesis_2022,kraemer_observation_2023,pineda_radiative_2025}) by a Gaussian signal on top of a Cherenkov-induced polynomial background, and a constant dark-count rate ($r_0$) (see~\cref{fig:spec_MgF2}). The observed VUV count rate $r$ was modeled as a function of the (time-dependent) grating position $x(t)$ using the following expression
\begin{multline}
    r(x(t)) = r_0 + (b_0e^{-\lambda_\mathrm{Ra}t}+b_1e^{-\lambda_\mathrm{Ac}t})\sum_{i=0}^2 c_ix^i \\+b_2\,f(t)\,e^{-\frac{(x-x_0)^2}{2\sigma^2}}\,.\label{eq: spectrum_description}
\end{multline}
The polynomial factors $c_i$ were extracted by fitting spectra obtained from $A=230$ implantations in the considered crystal, and are kept fixed. The strength of the contribution of $^{229}$Ra (resp. $^{229}$Ac) to the Cherenkov background is introduced by the a priori unknown parameter $b_0$ (resp. $b_1$), and the effect of the decay of the precursor nuclei during each scan was considered by applying the exponential corrections. The Gaussian factor is preceded by the amplitude, $b_2$, and a  factor $f(t)$ which is introduced to correct for the time behavior of the isomer population during the scan. For scans performed after the isomer has grown in, this correction factor is given by the exponential decay of $^{229}$Ac $f(t) = e^{-\lambda_\mathrm{Ac}t}$. To describe the grow-in character (which is heavily dependent on the a priori unknown quenching behavior), the spectra were first fitted without time correction, and the resulting peak amplitudes $b_2$ were fitted as a function of time with a second-degree polynomial. This polynomial was then used as correction factor in a subsequent fit of the spectra. This procedure was repeated until convergence, which was achieved after the first iteration.
\subsection{Quenching of the radiative decay}
To investigate the time behavior of the ensemble of isomers in the crystal, the fitted amplitudes ($b_2$) of the individual spectra are plotted as a function of time. The $\beta$ activity, originating from the precursor $A=229$ nuclei, is expected to induce quenching. As the total $\beta$ activity is time dependent, the observed quenching probability is expected to follow the time behavior of the precursors. In order to investigate the effect of $\alpha$ activity, $^{220}$Fr ions were implanted following the implantation of the $A=229$ chain.

The amount of quenching can be quantified by adapting the Bateman equation (that describes the pure radioactive decay sequence using the known half-lives of the involved isotopes) to include  a time-dependent decay constant $\lambda_Q(t)$. 
\begin{equation}
    \dv{N_\mathrm{iso}}{t} = \mathrm{RDF}\times\mathrm{BR}\times\lambda_\mathrm{Ac}N_\mathrm{Ac} - [\lambda_0+\lambda_Q(t)]\,N_\mathrm{iso}\,.\label{eq: time_behaviour model}
\end{equation}
Here, $\mathrm{BR}$ represents the total feeding probability of the $^{229}$Th isomer via the $\beta$ decay of $^{229}$Ac, and RDF is the radiative-decay fraction. We define RDF as the fraction of populated isomers that, in the absence of the activity-dependent quenching described by $\lambda_Q(t)$, reside in local structural-electronic configurations that remain radiatively active. Equivalently, these are configurations for which static states (local defect or excitonic) do not provide an efficient non-radiative decay channel at $E_{\mathrm{iso}}$. This definition absorbs time-independent and experimentally unresolved non-radiative losses into the overall normalization, rather than into the activity-dependent quenching parameters. The variable $N_{\mathrm{iso}}$ therefore denotes the number of isomers in the radiatively active subensemble. The decay constant $\lambda_0=\ln 2/t_{1/2}^{\mathrm{rad}}$ then describes the probability that an isomer ensemble will decay radiatively, while, conversely, $\lambda_Q(t)$ describes the probability of opening up the IC decay channel. Its time-dependence is determined by the total activity present in the crystal, for which a linear relationship is assumed, based on the work of Refs.~\cite{hiraki_Xray_2024,terhune2025photo}
\begin{equation}
    \lambda_Q(t)= q_\beta A_\beta + q_\alpha A_\alpha\,, \label{eq: quenching}
\end{equation}
where $q_\beta$ (resp. $q_\alpha$) describes the probability of a single $\beta$ (resp. $\alpha$) decay to open up the IC channel and $A_\beta$ and $A_\alpha$ the $\beta$ and $\alpha$ activity, respectively. With these definitions, the ratio $\lambda_0/\lambda_Q(t)$ can be interpreted as the instantaneous ratio for the isomer to decay radiatively compared to IC decay and other non-radiative decay channels.

As reported in Ref.~\cite{hiraki2025lasermossbauer}, the amount of laser-induced quenching (LIQ) of the radiative decay in CaF$_2$ is dependent on the local electronic environment of the isomer. In that laser M\"ossbauer spectroscopy experiment, performed on crystals doped with $^{229}$Th whereby the isomer was populated through laser excitation, different lattice configurations were identified. For all sites LIQ was observed, albeit with a different sensitivity. While the microscopic configurations in our implanted samples are not expected to coincide with those observed in Ref.~\cite{hiraki2025lasermossbauer}, it is still expected that implantation produces multiple configurations with different sensitivities to quenching. Even though our experiment does not allow us to distinguish the contribution of activity-induced quenching to all individual lattice sites, the time behavior of the radiative-decay rate in CaF$_2$ is better described when modeled with two fractions: one fraction of isomers which are susceptible to quenching induced by $\beta$ (resp. $\alpha$) activity with a probability density $q_\beta^{(1)}$ (resp. $q_\alpha^{(1)}$), and one fraction of isomers for which the quenching probability densities are negligible. Moreover, the MgF$_2$ crystal showed a much stronger sensitivity to quenching, resulting in datasets which could only be modeled with two fractions, both with a different nonzero quenching probability density. 

\subsection{Emission channeling.}
Emission channeling (EC) measurements were performed on the two epitaxial CaF$_2$/Si(111) thin films grown at $350$ \textdegree C and $850$ \textdegree C to determine the lattice sites occupied by implanted thorium. Following the procedure used for the CaF$_2$ single-crystal reference measurement in Ref.~\cite{kraemer_observation_2023}, the long-lived $^{229}$Th isotope was not used directly for EC. Instead, $^{231}$Th ($t_{1/2}=25.2~\si{\hour}$) was used as a chemical proxy and was populated by implantation of an $A=231$ radioactive ion beam, corresponding to the $^{231}$Fr$ \rightarrow ^{231}$Ra$ \rightarrow ^{231}$Ac $ \rightarrow ^{231}$Th decay chain. After a waiting time sufficient for the shorter-lived parent activities to decay, the angular emission patterns of electrons emitted in the decay of $^{231}$Th were recorded along the major crystallographic axes using a position-sensitive silicon detector and a three-axis goniometer. The measured two-dimensional emission patterns were fitted with simulated patterns calculated for thorium on high-symmetry CaF$_2$ lattice sites, including Ca-substitutional, F-substitutional, and interstitial sites, together with an isotropic random fraction. For the epitaxial films, the simulated patterns were additionally convoluted with a Gaussian angular broadening of standard deviation $\sigma$ to account for the loss of angular resolution expected due to the size of the beam spot on the implanted sample (for which an angular broadening around $\sigma=0.10^\circ$ is expected), but also mosaic/disorder broadening of the channeling features, as motivated by the known influence of crystal mosaicity on EC measurements in thin films~\cite{de2013influence}. The fit parameters reported below are the Ca-substitutional fraction $f_{\mathrm{S_{Ca}}}$, the root-mean-square displacement $u_1$ from the ideal Ca site, and the angular broadening parameter $\sigma$.
\section{Results}
\subsection{$\beta$-decay-induced quenching}
\renewcommand{\arraystretch}{1.3}
\begin{table*}[th]
\nolinenumbers
\caption{\label{tab:crystal_summary_beta_quenching} Summary of the $\beta$-decay quenching probabilities ($q_\beta$) in different crystals. The fraction $\xi$ corresponds to the fraction of isomers whose radiative decay is quenched with probability $q_\beta^{(1)}$. The CaF$_2$ datasets were not sensitive to a nonzero quenching probability in the other fraction. The data in CaF$_2$ UV could only constrain a single-fraction model. The \# datasets refers to the amount of time-behavior measurements have been performed in each sample. Whenever multiple datasets were obtained in a the same sample, they were jointly fitted using a hierarchical model.
}
\begin{tabular*}{\textwidth}{@{\extracolsep{\fill}}lllllll}
\hline
Crystal & \# datasets & $\xi$ & $q_\beta^{(1)}$ [Bq$^{-1}$] & $q_\beta^{(2)}$ [Bq$^{-1}$] & $\expval{q_\beta}$ [Bq$^{-1}$] & Relative RDF \\
\hline
CaF$_2$ 350 & 3 & $0.86\pm 0.03$ & $3.9^{+1.9}_{-1.5}\cdot 10^{-11}$& 0 & $3.4^{+1.6}_{-1.3}\cdot 10^{-11}$ & $0.56\pm 0.15$   \\
CaF$_2$ 850 & 2 & $0.63\pm 0.02$ & $1.0^{+0.6}_{-0.3}\cdot 10^{-10}$& 0 & $6.4^{+3.7}_{-2.1}\cdot 10^{-11}$ & $0.2 \pm 0.01$  \\
CaF$_2$ standard & 2 & $0.56\pm 0.04$ & $1.44^{+0.71}_{-0.49}\cdot 10^{-10}$& 0 & $8.0^{+3.6}_{-2.4}\cdot 10^{-11}$ & $1.2\pm 0.2$  \\
CaF$_2$ UV & 1 & 1 & $6.5^{+1.6}_{-1.4}\cdot 10^{-12}$& 0& $6.5^{+1.6}_{-1.4}\cdot 10^{-12}$& $1$   \\
MgF$_2$ & 3 & $0.62^{+0.06}_{-0.05}$ & $1.3^{+0.6}_{-0.4}\cdot 10^{-9}$& $5.1^{+1.5}_{-1.7}\cdot 10^{-11}$& $8.2^{+3.3}_{-1.9}\cdot 10^{-10}$ & $1.4 \pm 0.4$   \\
\hline
\end{tabular*}
\linenumbers
\end{table*}
\renewcommand{\arraystretch}{1.0}
The time behavior of the radiative decay was studied in different types of CaF$_2$ crystals and in MgF$_2$. The model used for describing the time behavior (\cref{eq: time_behaviour model}), requires the knowledge of the (unquenched) radiative decay constant $\lambda_0=\ln 2/t_{1/2}^{\mathrm{rad}}$.  Due to the Purcell effect, a medium with refractive index $n$ leads to a scaling of the radiative half-life by a factor of $n^{-3}$ compared to the half-life in vacuum~\cite{Tkalya2015_radLifetime_PRC}. For CaF$_2$, a radiative half-life of $444\pm 3$~s (as reported by Ref.~\cite{zhang_frequency_2024}) is assumed. Since the radiative half-life in MgF$_2$ has not been measured precisely, an estimated value is used based on the $n^{-3}$ relationship, assuming refractive indices at $148.5\,\si{\nano\meter}$ of $n_{\mathrm{CaF}_2}\approx1.582$~\cite{li1980refractive} and $n_{\mathrm{MgF}_2}\approx1.497$~\cite{li1980refractive}. This results in a radiative halflife of the isomer in MgF$_2$ of 523~s, which lies on the lower end of the earlier measurement of $670\pm 102\,\si{\second}$~\cite{kraemer_observation_2023}.

Four model parameters need to be inferred from the data: the two $\beta$-decay quenching probabilities ($q_\beta^{(1)}$ and $q_\beta^{(2)}$), the fraction $\xi$ of the isomers whose probability to decay radiatively is quenched with $q_\beta^{(1)}$, and an overall scaling parameter $\eta$, taking into account the total beta-decay feeding of $^{229}$Ac towards the isomer, the radiative decay fraction, and the geometrical and instrumental efficiency of the spectrometer. The parameters were determined within a Bayesian framework, and the posterior distributions were sampled using an affine-invariant Markov Chain Monte Carlo (MCMC) sampler. 

When multiple datasets were obtained from the same crystal, they are fitted jointly: dataset-dependent quenching probabilities $q_{\beta,i}^{(1)}$ and $q_{\beta,i}^{(2)}$ are introduced for each dataset $i$, and they are assigned Gaussian hierarchical priors of the form
\[q_{\beta,i}^{(k)}\sim\mathcal{N}(\mu_q^{(k)}, \tau^{(k)})\quad k\in \{1,2\}\,,\]
where $\mu_q^{(k)}$ represents the common mean and $\tau^{(k)}$ the dataset-to-dataset scatter. This allows for systematic differences between datasets while constraining them through a shared underlying value.

\begin{figure*}[htp]
    \centering
    \includegraphics[width=.5\linewidth]{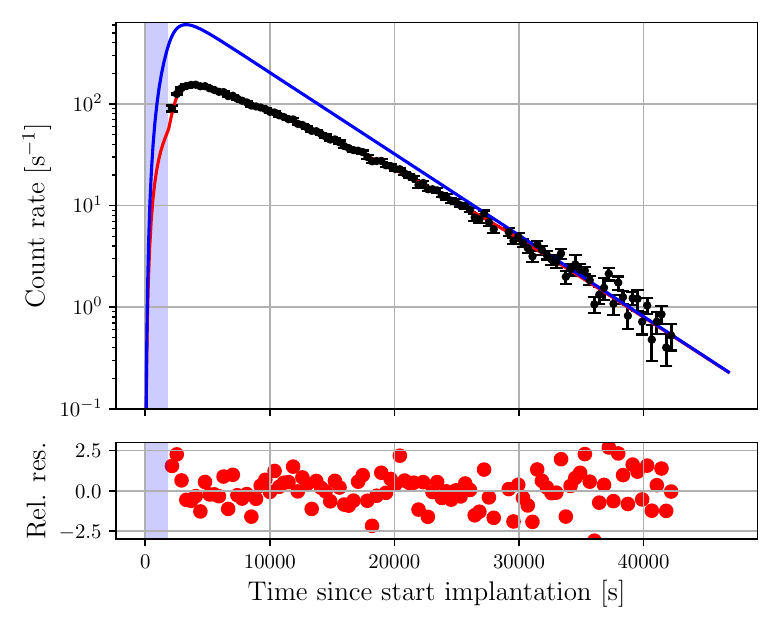}
    \includegraphics[width=.5\linewidth]{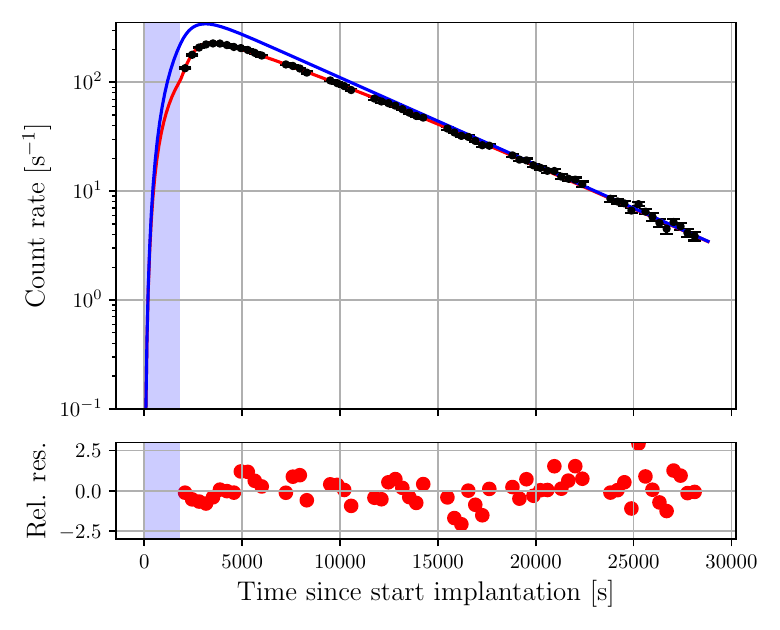}
    \caption{Time behavior of the radiative-decay rate in MgF$_2$ (left) and CaF$_2$ 350 (right), assuming a half-life of 523 s (resp. 444 s) for MgF$_2$ (resp. CaF$_2$). The blue-shaded region illustrates the $A=229$ implantation period. In red: the posterior predictive median of the data from the MCMC analysis. In blue: the expected time behavior assuming no quenching was involved. Bottom: residuals scaled to the uncertainties on the data points.}
    \label{fig:PPM_beta}
\end{figure*}

The results of this analysis are summarized in \cref{tab:crystal_summary_beta_quenching}. Only the datasets collected in MgF$_2$ could constrain a full two-fraction model with two nonzero $\beta$-quenching probability densities. The resulting posterior predictive median for the time behavior curve in MgF$_2$ is plotted (in red) against the data in the left panel in \cref{fig:PPM_beta}. For comparison, a standard Bateman curve (without quenching), assuming the same overall efficiency $\eta$, is plotted in blue.  The CaF$_2$ crystals showed a weaker sensitivity to $\beta$-decay-induced quenching and datasets obtained in CaF$_2$ 350, CaF$_2$ 850 and the standard-grade CaF$_2$ bulk crystal could only constrain a two-fraction model where a single fraction of the isomers is sensitive to quenching. The CaF$_2$ UV grade crystal showed such weak sensitivity to quenching that, with the observed count rates, the data could only constrain a single-fraction model.

Using the results of this analysis, the average quenching probability density $\expval{q_\beta}=\xi\,q_\beta^{(1)} +(1-\xi)\,q_\beta^{(2)} $ is calculated for each sample. Relative to CaF$_2$ UV-grade, the global sensitivity of radiative decay to $\beta$-decay-induced quenching is higher by a factor of $12^{+7}_{-5}$ in standard-grade CaF$_2$ (Thorlabs) and by $129^{+63}_{-39}$ in MgF$_2$.

Additionally, the overall scaling parameter $\eta$ can be used to estimate a relative RDF. As discussed in Ref.~\cite{kraemer_observation_2023}, because of limited knowledge of the absolute feeding of the isomer in the $\beta$ decay of $^{229}$Ac, only a lower limit for RDF can be deduced. However, since the geometrical efficiency is dataset-dependent, $\eta$ was considered a nuisance parameter, and a posterior distribution was determined for each dataset. The medians of these posterior distributions were corrected for the geometrical efficiency of the VUV spectrometer, which depends on the sample–slit distance and slit size. The corrected values were then combined, for each sample, using an arithmetic mean and standard deviation to obtain a conservative estimate of the expectation value and its systematic spread. The results are reported relative to the value obtained in the CaF$_2$ UV grade sample (where only one dataset was collected), hereby eliminating the uncertainties related to the VUV spectrometer's intrinsic efficiency. Contrarily to Ref.~\cite{pineda_radiative_2025}, where the RDF was determined from a spectrum at the apex of the time-behavior curve (where $\beta$-decay-induced quenching does play a significant role, but was not accounted for), the current results successfully decouple the RDF from the quenching probabilities $q^{(k)}_\beta$. It is noted that the $\beta$-decay-induced quenching probabilities obtained from the standard and UV-grade CaF$_2$ samples, and the MgF$_2$ sample span up to two orders of magnitude, while the relative RDF values are, within uncertainties, consistent. 

It should be noted that the conclusions drawn from the data obtained in MgF$_2$ depend on the choice of the radiative half-life. The value adopted here relies on $n^{-3}$ scaling of the well-measured half-life~\cite{li1980refractive} in CaF$_2$ and should therefore be regarded as an estimate. In particular, the refractive index of MgF$_2$ at $148.5\,\si{\nano\meter}$ is extrapolated, and the use of the $n^{-3}$ scaling might be limited for implanted ions located only a few nanometers below the crystal surface. Allowing the MgF$_2$ radiative half-life to vary as an additional free parameter in the MCMC analysis results in a lower preferred value , but introduces strong parameter correlations. Since a shorter assumed radiative half-life leads to larger inferred quenching probabilities, the conclusions regarding the remarkably strong quenching in MgF$_2$ compared to CaF$_2$ as  obtained with the value of $523\,\si{\second}$ are conservative.

\subsection{$\alpha$-decay-induced quenching}

\begin{figure}[ht]
    \centering
    \includegraphics[width=\linewidth]{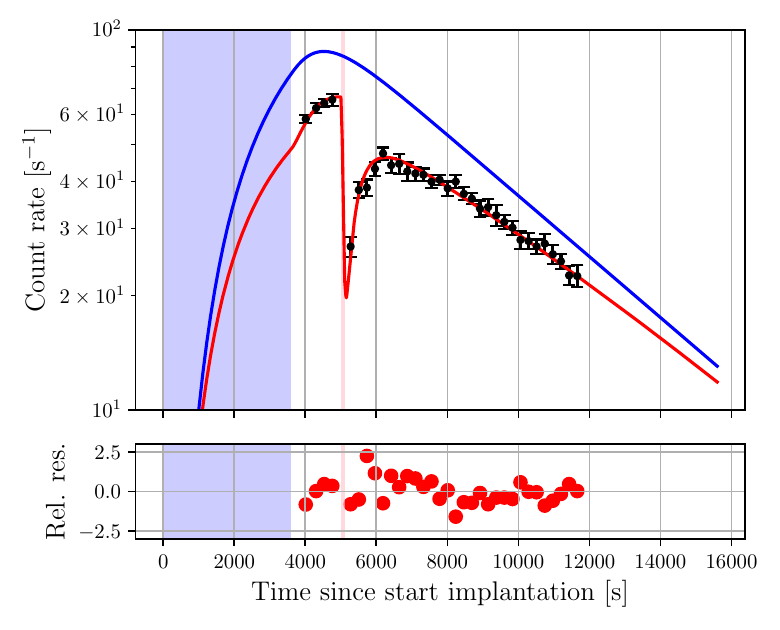}
    \caption{Effect of $\alpha$ radiation on the time behavior of the radiative-decay rate in CaF$_2$. The blue-shaded region indicates the 1-hour $A=229$ implantation, while the pink-shaded region indicates the 2-minute $A=220$ implantation. The latter beam consisted mainly of the $\alpha$ emitter $^{220}$Fr (t$_{1/2}$ = 27.4 $\si{\second}$). In red: the posterior predictive median of the data from the MCMC analysis. In blue: the expected time behavior assuming no quenching was involved. Bottom: residuals scaled to the uncertainties on the data points.}
    \label{fig:PPM_alpha}
\end{figure}

A CaF$_2$ 350 crystal was used to investigate $\alpha$-decay induced quenching. An $A=229$ beam was implanted for 1 hour. After performing a few scans to determine the radiative-decay-signal height (which is quenched by the $\beta$ decay of the $A=229$ precursors) the crystal was positioned back in the implantation position by rotating the sample wheel. A two-minute $A=220$ implantation took place $1410\,\si{\second}$ after the end of the $A=229$ implantation (\cref{fig:PPM_alpha}) and after positioning the crystal back in front of the slits, wavelength scanning was resumed at $1605\,\si{\second}$ after the end of the $A=229$ implantation. The data evidently show that the VUV signal was abruptly quenched by the short-lived members of the $^{220}$Fr$\rightarrow ^{216}$At$\rightarrow ^{212}$Bi $\alpha$ decay chain, which quickly reaches transient equilibrium with the longer-lived $^{212}$Bi. Since the isomeric state is still being populated by the $\beta$ decay of $^{229}$Ac, the VUV signal grows in on a time-scale determined by the radiative half-life of $^{229\mathrm{m}}$Th in CaF$_2$, while still being quenched by the remaining $\alpha$ and $\beta$ activities originating from the $^{212}$Bi and $^{229}$Ac decay chains.

As for the determination of the $\beta$-decay-induced quenching probabilities in CaF$_2$ 350, a two-fraction model where only one of the fractions has a nonzero quenching probability density could be constrained. Similarly, the posterior distributions for $q_\beta$, $q_\alpha$ and $\xi$ where inferred with medians $q^{(1)}_\beta = 1.8\pm0.5\cdot10^{-11}~\mathrm{Bq}^{-1}$, $q^{(1)}_\alpha = 7\pm3\cdot10^{-11}~\mathrm{Bq}^{-1}$, and $\xi = 0.81^{+0.03}_{-0.02}$. No prior knowledge on $q_\beta^{(1)}$ or $\xi$ from earlier experiments was assumed and the obtained value for $\xi$ and $q^{(1)}_\beta$ are, within uncertainty, consistent with the values reported in~\cref{tab:crystal_summary_beta_quenching}.   Using the posterior distributions for $q^{(1)}_\alpha$ and $q^{(1)}_\beta$, the relative $\alpha$-quenching probability density can be determined through the ratio $q^{(1)}_\alpha/q^{(1)}_\beta$, delivering a value of $4.0\pm 0.7$. 

\subsection{Emission channeling}
\begin{figure}[t]
    \centering
    \includegraphics[width=\linewidth]{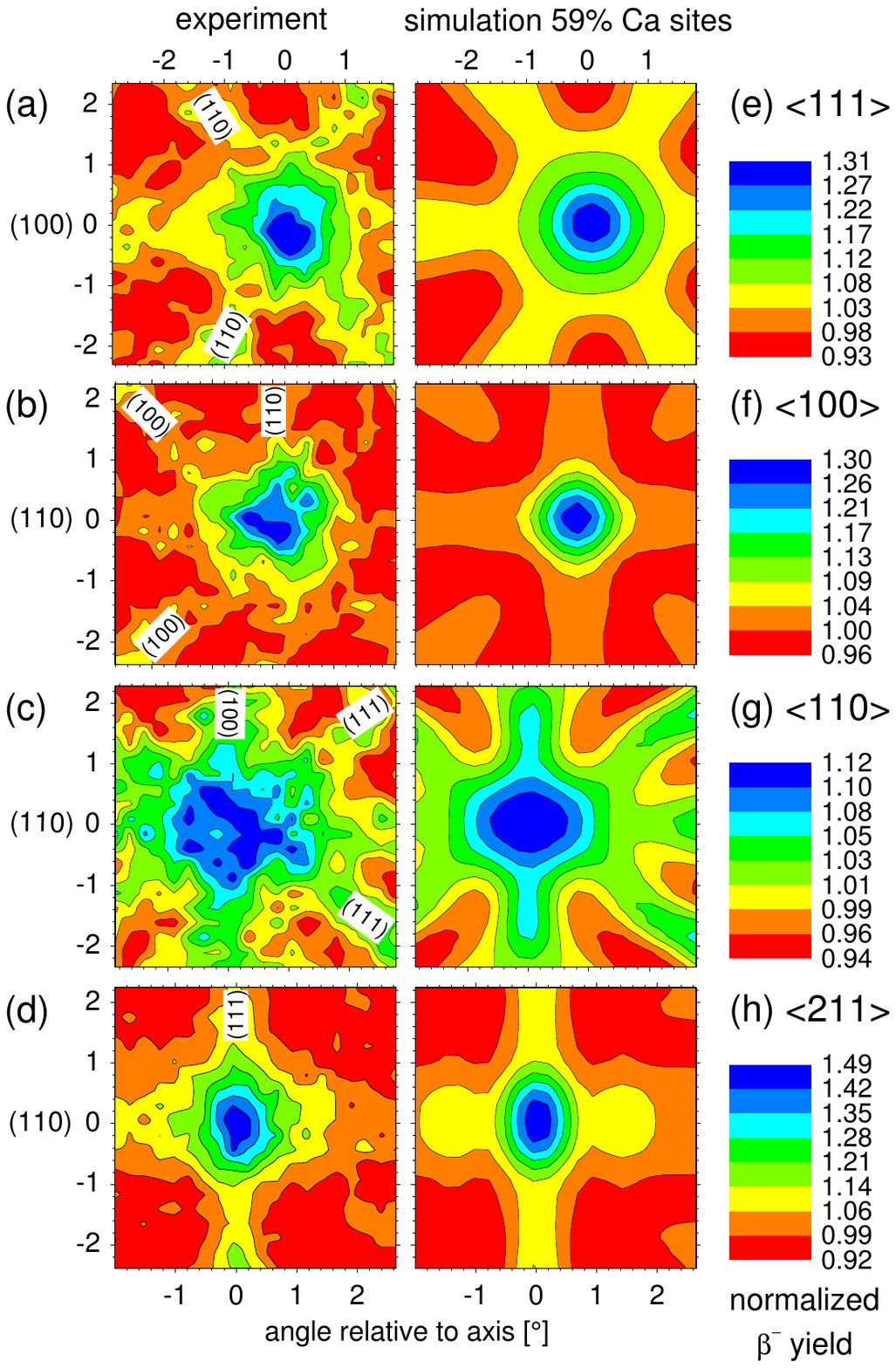}
    \caption{Experimental and best-fit simulated emission channeling patterns of electrons emitted by $^{231}$Th in the CaF$_2$ 350 film, measured in the as-implanted
    state along the indicated crystallographic directions. The best fit corresponds to Th on Ca-substitutional sites, including an isotropic random fraction and Gaussian angular broadening. The fit yields
    $f_{\mathrm{S}_{\mathrm{Ca}}}=59(3)\%$, $u_1=0.15~\mathrm{\AA}$, and
    $\sigma=0.25^\circ$.}
    \label{fig:ec_caf2_350}
\end{figure}

\begin{figure}[t]
    \centering
    \includegraphics[width=\linewidth]{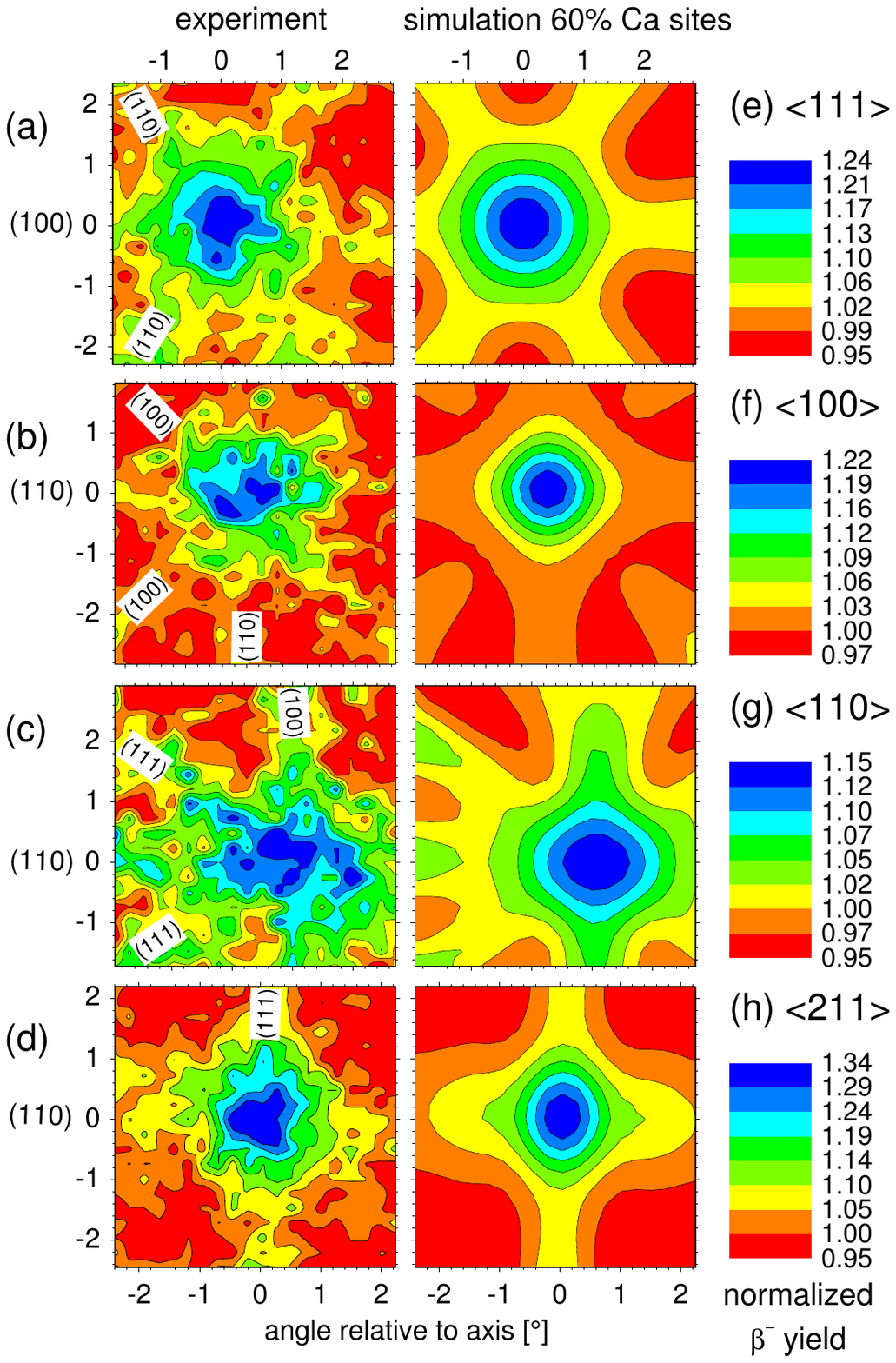}
    \caption{Experimental and best-fit simulated emission channeling patterns of electrons emitted by $^{231}$Th in the CaF$_2$ 850 film, measured in the as-implanted state along the indicated crystallographic directions. The best fit corresponds
    to Th on Ca-substitutional sites, including an isotropic random fraction and Gaussian angular broadening. The fit yields
    $f_{\mathrm{S}_{\mathrm{Ca}}}=60(9)\%$, $u_1=0.13~\mathrm{\AA}$, and
    $\sigma=0.40^\circ$.}
    \label{fig:ec_caf2_850}
\end{figure}

To compare the local incorporation of thorium in the two CaF$_2$ thin films with the single-crystal CaF$_2$ reference of Ref.~\cite{kraemer_observation_2023}, analogous EC measurements were performed on the CaF$_2$ 350 and CaF$_2$ 850 samples. The best-fit Ca-substitutional fractions, root-mean-square displacements, and angular broadening parameters are summarized in \cref{tab:EC_films}. The experimental EC patterns and corresponding best-fit simulations for the as-implanted CaF$_2$ 350 and CaF$_2$ 850 films are shown in Figs.~\ref{fig:ec_caf2_350} and~\ref{fig:ec_caf2_850}, respectively.

\begin{table}[h!]
    \centering
    \caption{Emission channeling fit results for $^{231}$Th implanted in the
    CaF$_2$ thin films. Here $f_{\mathrm{S_{Ca}}}$ denotes the fraction of Th
    atoms occupying Ca-substitutional sites, $u_1$ is the root-mean-square
    displacement from the ideal Ca site, and $\sigma$ is the total Gaussian angular broadening, including the experimental resolution contribution and additional mosaic/disorder broadening of the channeling patterns. The random fraction is defined as
    $f_{\mathrm{rand}}=1-f_{\mathrm{S_{Ca}}}$.}
    \label{tab:EC_films}
    \begin{ruledtabular}
        \begin{tabular}{lcccc}
        Sample &
        $f_{\mathrm{S_{Ca}}}$ &
        $u_1$ (\AA) &
        $\sigma$ ($^\circ$) &
        $f_{\mathrm{rand}}$ \\
        \hline
        CaF$_2$ 350 & $(59\pm3)\%$ & $0.15$ & $0.25$ & $(41\pm3)\%$ \\
        CaF$_2$ 850 & $(60\pm9)\%$ & $0.13$ & $0.40$ & $(40\pm9)\%$ \\
        \end{tabular}
    \end{ruledtabular}
\end{table}

The two films show the same Ca-substitutional Th fraction within uncertainty:
$f_{\mathrm{S_{Ca}}}=59(3)\%$ for CaF$_2$ 350 and $60(9)\%$ for CaF$_2$ 850.
The best-fit displacements from the Ca site, $u_1=0.15\,\si{\angstrom}$ and
$0.13\,\si{\angstrom}$, are also comparable; the difference is within the sensitivity of the present EC analysis. The larger angular broadening parameter in the CaF$_2$ 850 film, $\sigma=0.40^\circ$, compared with $\sigma=0.25^\circ$ for CaF$_2$ 350, indicates a stronger broadening and degradation of the channeling features, consistent with a more disordered or more mosaic film~\cite{de2013influence}.

The fitted Ca-substitutional fractions in both films are well below unity. This contrasts with the single-crystal CaF$_2$ EC data of Ref.~\cite{kraemer_observation_2023}, for which the Ca-substitutional fraction is close to $100\%$ after applying the same correction for secondary-electron contributions. The remaining non-substitutional contribution in the present EC
data is the so-called random fraction, corresponding to Th atoms in low-symmetry local environments that give rise to a nearly isotropic EC pattern. These random fractions are large and similar in the two films, $f_{\mathrm{rand}}\simeq40\%$. The sizable random fractions provide a natural microscopic connection to the non-radiative fractions inferred from the VUV spectroscopy results. Suppression of IC is controlled by the local electronic spectral density coupled to the Th nucleus at $E_{\mathrm{iso}}$, rather than by the ideal host band gap alone. Suitably charge-compensated, substitutional Th in well ordered crystalline CaF$_2$ is expected to have little such spectral weight and therefore to remain radiatively active. In highly disordered CaF$_2$, however, the local environment can contain defect, impurity, or carrier-trapping states inside the nominal host gap. These states may provide occupied initial states, empty final states, or both, and can thereby enable IC. 
Thus, Th atoms in the EC random fraction are likely to contribute to the non-radiative component of the isomer ensemble. At the same time, the EC random fractions alone do not fully explain the difference between the two films. The random fractions are similar in CaF$_2$ 350 and CaF$_2$ 850, whereas the VUV analysis gives relative radiative-decay fractions of approximately $0.56$ and $0.20$, respectively, corresponding to missing radiative fractions of order $50\%$ and $80\%$ in the present normalization. For CaF$_2$ 350, the EC random fraction can therefore account for most of the inferred non-radiative fraction. In CaF$_2$ 850, however, the much larger non-radiative fraction suggests that a significant subset of Th atoms classified by EC as Ca-substitutional still resides in defective local CaF$_2$ environments that allow internal conversion. This interpretation is consistent with the larger $\sigma$ value for the CaF$_2$ 850 film, which indicates stronger mosaic/disorder broadening and hence a less ideal local crystalline environment.

\section{Discussion}
\begin{figure*}[htp]
\includegraphics[width =\textwidth]{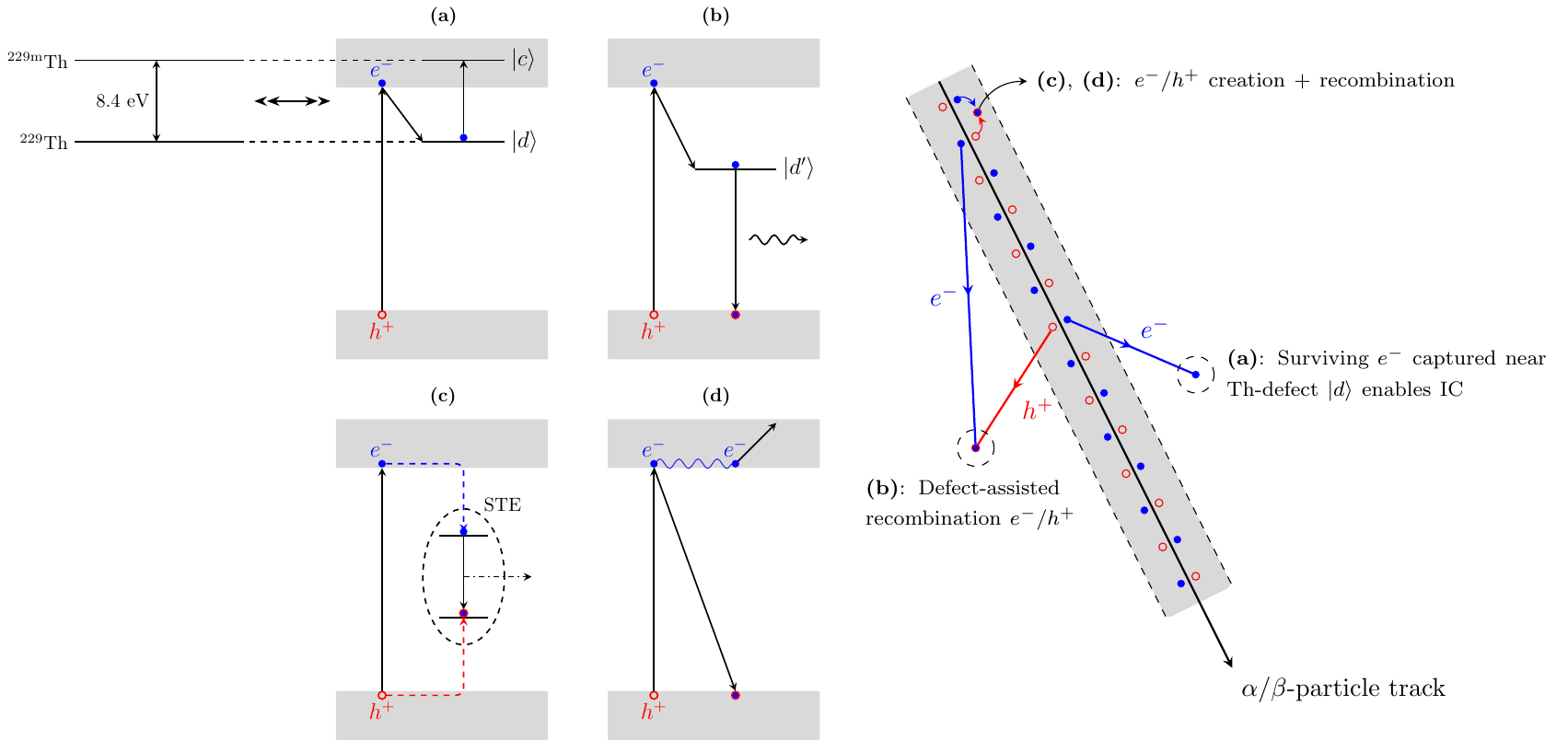}
    \caption{\textbf{Left}: Conceptual illustration of radiation-induced quenching based on the creation of electron-hole pairs. (a) In the simplest realization of our proposed mechanism, an electron produced by the $\alpha$ or $\beta$ ionization cascade, or released from a trap, is captured by a Th defect state $\ket{d}$. The resulting occupied local state can participate in an energy-conserving IC transition through the coupling between the nuclear and electronic degrees of freedom~\cite{morgan2024ICtheory}. A related possibility is STE- or trapped-carrier-assisted quenching, in which a local excitation formed near the Th defect complex modifies the charge configuration or transfers energy in a way that enables IC. The other diagrams show the competing electron-hole recombination mechanisms. (b) Defect-assisted carrier trapping and recombination. (c) STE formation followed by radiative decay, non-radiative decay, or defect/color-center formation. (d) Auger-like or other high-density many-body recombination. \textbf{Right}: Diagram illustrating the electron-hole pair creation along the particle track in real space, and highlighting the possible recombination mechanisms. The gray-shaded area indicates the radius of the ionization track, which is comparable for $\alpha$ and $\beta$ radiation~\cite{cang2020ionization}.}
    \label{fig:quench_recomb_mechanism}
\end{figure*}
Measuring the time behavior of the radiative decay of $^{229\mathrm{m}}$Th in different solid-state hosts allowed to quantify the effect of quenching induced by radioactive decay. Following the carrier-mediated quenching picture proposed in Ref.~\cite{guan_xray_2026}, we interpret the activity-dependent quenching as arising from electronic excitations generated by the decay radiation, which survive prompt recombination and become captured, stored, or localized near Th defects, thereby enabling non-radiative de-excitation of the isomer (\cref{fig:quench_recomb_mechanism}, panel \textbf{(a)}). 

In the simplest realization, these excitations are mobile or weakly localized electrons produced along the $\alpha$ and $\beta$ ionization tracks and subsequently captured by a Th-associated trap, thereby opening an IC channel (\cref{fig:quench_recomb_mechanism}, right panel). In the following, we use this electron-capture process as the leading microscopic picture. However, the fitted $q_\alpha$ and $q_\beta$ values are effective per-decay probabilities and may also include delayed release from color centers, hole-assisted charge-state changes, or self-trapped exciton (STE) related excitations located near a Th defect. An important observation is the factor of $ 4.0\pm 0.7$ difference in the quenching probabilities for $\alpha$ and $\beta$ decay observed within the CaF$_2$ 350 sample. Since $\alpha$ particles in the MeV-energy range have stopping powers which are three orders of magnitude larger than $\beta$ particles in the $100\,\si{\kilo\electronvolt}$-energy range, $\alpha$ particles generate orders of magnitude more electron-hole pairs per unit length than $\beta$ particles do. However, the relevant quantity for quenching is not simply the initial pair yield, but the density of mobile electrons that survive the rapid recombination processes within the track. To understand the difference in quenching probability density between $\alpha$ and $\beta$ radiation, it is necessary to consider the competition between electron-hole pair generation and recombination.  

A commonly used phenomenological description for the recombination rate per unit volume $R_\mathrm{rec}$ is the ABC model~\cite{hopkins2017abc}
\begin{equation}
    R_\mathrm{rec} = An+Bn^2+Cn^3\,, \label{eq:ABC}
\end{equation}
where it is assumed, to first approximation, that the ionization cascade creates electrons and holes in equal numbers, such that the carrier density $n_e\approx n_h\approx n$. The three terms in \cref{eq:ABC} represent distinct recombination mechanisms, which are illustrated in panels \textbf{(b)}, \textbf{(c)} and \textbf{(d)} of \cref{fig:quench_recomb_mechanism}. The first term, $An$, represents defect-assisted Shockley-Read-Hall recombination. It scales linearly with the charge-carrier density and with the defect density. The $Bn^2$ term represents bimolecular electron-hole recombination. In wide-band-gap fluorides this channel should be understood broadly: the initially created electron-hole pair can form an exciton and subsequently self-trap, giving rise to STE radioluminescence, non-radiative STE decay, or defect formation. The $Cn^3$ term describes Auger-like and other high-density many-body recombination processes. Therefore, while the electron-hole creation rate scales linearly with the stopping power, so does (to first approximation) the recombination rate per electron-hole pair $R_\mathrm{rec}/n$. Thus, increasing stopping powers has two opposing effects: while it increases the electron-hole pairs initially created, it decreases the time those carriers survive before recombining. This naturally leads to the quenching mechanism we propose. The initial ionization track has a limited radius of a few nanometers, whereas secondary (mobile) electrons can travel well beyond that dense core, reaching a radial distribution maximum around $30\,\si{\nano\meter}$ in CaF$_2$ and exhibiting a tail that extends out to $100$-$200\,\si{\nano\meter}$~\cite{wang2012monte}. The population of mobile electrons which survive recombination can diffuse to thorium sites, where their capture enables IC. This surviving excitation population should therefore not necessarily be interpreted as a population of free conduction-band electrons persisting over the full quenching time scale. A related possibility is STE-assisted quenching near a Th-associated defect complex: an STE, or an STE-derived trapped excitation, formed near the Th defect could transfer energy or modify the local charge configuration in a way that enables IC. The present data cannot distinguish whether the active species is a free or weakly localized electron, a trapped electron released from a color center, or an STE-derived excitation. The extracted $q_\alpha$ and $q_\beta$ should therefore be regarded as effective probabilities that include carrier creation, rapid recombination, self-trapping, delayed trap release, and capture or localization near Th-associated defects. The observation of multiple $^{229\mathrm{m}}$Th fractions with different quenching probabilities can equally be understood within this picture. The recent laser-M\"ossbauer spectroscopy results reported in Ref.~\cite{hiraki2025lasermossbauer} revealed four distinct $^{229}$Th configurations in CaF$_2$ that remain radiatively active in the absence of an additional captured electron. In this situation, at least two factors can influence the effective quenching probability density $q_\beta^{(k)}$. First, different thorium configurations will generally have different trapping  cross sections, since the capture probability depends on the local defect potential, the spatial extent and energy of the trap state, and the surrounding charge-compensation structure. Second, once an electron is trapped, the corresponding IC probability need not be universal. As discussed in~Ref.\cite{morgan2024ICtheory}, the IC rate is not determined only by the energy of a defect level. It also depends on the local nuclear-electronic coupling, which is controlled by the electronic wave-function amplitude near the Th nucleus and by the Th-projected character of the participating defect-state components. Different Th configurations can therefore differ both in how efficiently they capture or localize radiation-generated carriers and in how efficiently the resulting charged or excited defect complex couples to the nuclear isomer.

This picture further allows to explain the observed variation in $\beta$-decay quenching probabilities  within different samples. The spectra from the $A=230$ implantations indicate the presence of color centers in the more strongly quenching standard-grade CaF$_2$ bulk crystal from Thorlabs. These were not observed in the CaF$_2$ UV grade sample from Crystran, which showed the lowest sensitivity to quenching. Furthermore, compared to high-quality single crystals, thin films are expected to have a higher density of defects, which aligns with the higher observed sensitivity to quenching compared to CaF$_2$ UV. This trend that quenching is stronger in the more defective (non-UV-grade and thin films) than in the least defective (UV grade) CaF$_2$ suggests that the additional defects promote charge separation or temporary charge storage, instead of acting as prompt nonradiative recombination centers. In doing so, the defects increase the population of electrons that survive prompt electron-hole recombination and remain available for subsequent capture at thorium-related sites. This interpretation is also consistent with the quenching model of Ref.~\cite{guan_xray_2026}, in which crystal defects in CaF$_2$ act as charge traps rather than prompt recombination centers. In this picture, the afterglow observed in Ref.~\cite{guan_xray_2026} reflects delayed recombination of trapped electrons and holes, while trapped electrons can be released again and remain available either for recombination or for subsequent capture at thorium-related sites, thereby contributing to isomer quenching. The observation of a color-center emission band in our most defective sample is likewise consistent with the presence of defect-related trapping states, supporting the view that these defects participate in charge storage and delayed release rather than acting only as prompt recombination centers. 

A detailed interpretation of the stronger observed quenching in MgF$_2$ is difficult at present, since the effective quenching probability density can depend on several host-specific microscopic processes. Compared with CaF$_2$, MgF$_2$ may differ in the initial electron-hole-pair yield and thermalization, the probability and time scale for STE formation, phonon-assisted non-radiative relaxation of STEs or trapped excitations, the density and energy spectrum of charge traps, and the capture cross section of Th-associated defect complexes. These processes can have competing effects: rapid non-radiative relaxation may reduce the number of free carriers, but it may also populate trapped or STE-derived excitations that survive long enough to interact with Th-associated complexes. The larger quenching probability density observed in MgF$_2$ should therefore be interpreted as a larger effective probability for radiation-generated electronic excitation to enable IC, rather than as evidence for a single microscopic parameter.

In addition to supporting electron trapping at impurities as the microscopic origin of the quenching mechanism, our observations disfavor alternative mechanisms. First, a photon-mediated process, like scintillation or Cherenkov emission, is inconsistent with the relatively small $q^{(1)}_\alpha/q^{(1)}_\beta$ ratio. Scintillation should scale directly with the photon yield produced by the energy deposited along the particle track~\cite{cang2020ionization,kawaguchi2018scintillation,li2020study}, which would imply a significantly stronger difference between $\alpha$- and $\beta$-decay-induced quenching.  Moreover, a Cherenkov-based mechanism would predict enhanced quenching in the bulk crystals compared to the thin films due to longer $\beta$-track lengths, while, contrarily, the observed quenching probabilities of the thin films lie in between those of the bulk crystals. Second, mechanisms governed directly by prompt lattice heating or by permanent irradiation-induced structural rearrangements ~\cite{schaden_LIQ_2025} are also disfavored as the dominant direct quenching pathway. Such processes should scale more directly with deposited energy, leading to a much larger $q^{(1)}_\alpha/q^{(1)}_\beta$ ratio. This argument does not exclude phonon-assisted relaxation within a carrier-, trap-, or STE-mediated pathway; it only disfavors a quenching mechanism whose rate is controlled directly by local heating or permanent displacement damage.

The results of $\alpha$- and $\beta$-decay-induced quenching probabilities allow to estimate the magnitude of quenching due to the decay of ground-state $^{229}$Th in typical samples used in recent work~\cite{tiedau_laser_2024, zhang_frequency_2024,ooi_reproducibilty_2026}. For a thorium density of $5\cdot10^{18}\,\si{\per\centi\meter\cubed}$ in a millimeter-scale crystal, the total activity of ground-state $^{229}$Th is on the order of $25\,\si{\kilo\becquerel}$, which, under approximate secular equilibrium of the decay chain, corresponds to a combined $\alpha$- and $\beta$-decay rate. Using the measured values of  $q^{(1)}_\alpha$ and $q^{(1)}_\beta$ in CaF$_2$ 350, this yields an overall quenching effect size of order $10^{-5}$. 
\section{Conclusion}
In this work, we have demonstrated that the radiative decay of an ensemble of $^{229\mathrm{m}}$Th nuclei embedded in solid-state hosts can be quenched by $\beta$ and $\alpha$ radiation. The isomer was populated through the $\beta$ decay of $^{229}$Ac, and the time behavior of its radiative decay signal was monitored using VUV spectroscopy. Based on this, we introduced a model which allows to quantify the relative strength of $\alpha$- and $\beta$-decay-induced quenching in the form of quenching probabilities.

Overall, our results from MgF$_2$ and CaF$_2$ indicate the presence of at least two fractions of isomers residing in different microscopic configurations with different quenching probabilities. This supports the recent observation in laser-M\"ossbauer spectroscopy studies of multiple lattice-site configurations that remain radiatively active in the absence of an additional captured electron \cite{hiraki2025lasermossbauer}. The observed variation between different types of CaF$_2$ crystals further highlights the relevance of crystal quality and defect density in determining the overall quenching probability density.

Our methodology also allowed to successfully decouple the effect of the quenching induced by the radioactive decay of the precursors from the radiative-decay fraction. This revealed that the relative RDF values obtained in the bulk crystals are, within uncertainties, mutually consistent. This includes crystals of different host materials and different defect densities. In contrast, the quenching probabilities show a clear dependence on crystal quality, suggesting that, while crystal defects strongly influence the sensitivity to radiation-induced quenching, they do not substantially modify the RDF near the crystal surface.

The quenching probabilities are consistent with a trap-mediated microscopic mechanism. In the simplest realization, radiation-generated electrons that survive prompt recombination are captured by Th defects, thereby opening an IC channel. More generally, the measured $q_\alpha$ and $q_\beta$ values are effective probabilities that may also include color-center trapping and release, STE formation and decay, and host-dependent trap spectra. Within this broader picture, with $q_\alpha$ being only four times larger than $q_\beta$, the difference between $\alpha$- and $\beta$-decay-induced quenching remains small. This can be explained by the competition between electron-hole pair generation and rapid recombination or self-trapping, which limits the surviving excitation population at high stopping power.

Finally, an estimate based on the measured quenching probabilities, shows that radiation-induced quenching from the radioactivity of $^{229}$Th doped in CaF$_2$ remains negligible for typical thorium densities used in current experiments. These results provide relevant information on the role of radiation-induced processes in solid-state hosts which are suitable for the development of a nuclear clock. They also support the scenario of quenching mediated by charge carriers as the dominant mechanism, with electron capture at Th defects as the simplest microscopic realization. Future studies will apply the methodology from this work to additional crystalline host materials, and the effect of temperature on radiation-induced quenching will be studied.

\begin{acknowledgments}
We acknowledge the support of the ISOLDE Collaboration and technical teams. Additionally, UW and JGC acknowledge support from FCT- Portugal, 2024.00223.CERN funded by the PRR, RE-C06-i06.m02, through EMRP and FCT (\url{https://doi.org/10.54499/2024.00223.CERN}), SB is supported by a fellowship from Research Foundation Flanders (FWO) under grant agreement nr. 1167324N and SK is supported by a fellowship from Research Foundation Flanders under grant agreement nr. 12A1026N. This work is supported in part by Research Foundation Flanders grants G078624N and I001323N, the FWO-FNRS project 40007501 under the EOS program, the Special Research Fund (BOF KU Leuven C14/22/104), the European Union’s Horizon Europe Framework research and innovation programme under grant agreement no. 101057511 (EURO-LABS), the European Union’s Horizon 2020 research and innovation programme under the Marie Skłodowska-Curie grant agreement no. 101026762 and 861198 (LISA), the European Research Council (ERC ThoriumNuclearClock, grant agreement no. 856415), the European Union’s Horizon 2020 research and innovation programme under grant agreement no. 819957, the Romanian IFA grant CERN/ISOLDE, the Nucleu project no. PN 23 21 01 02, the Bavaria California Technology Center (BaCaTeC 7 [2019-2]), and the German Federal Ministry of Research, Technology and Space (BMFTR) (Quantum Futur II Grant ``NuQuant'', grant agreement no. 13N16295A).

\end{acknowledgments}


\bibliography{apssamp}

@PREAMBLE{
 "\providecommand{\noopsort}[1]{}" 
 # "\providecommand{\singleletter}[1]{#1}%" 
}

@article{kraemer_observation_2023,
	title = {Observation of the radiative decay of the $^{229}${T}h nuclear clock isomer},
	volume = {617},
	pages = {706--710},
	number = {7962},
	journal = {Nature},
	author = {Kraemer, Sandro and Moens, Janni and Athanasakis-Kaklamanakis, Michail and Bara, Silvia and Beeks, Kjeld and Chhetri, Premaditya and Chrysalidis, Katerina and Claessens, Arno and Cocolios, Thomas E. and Correia, João G. M. and Witte, Hilde De and Ferrer, Rafael and Geldhof, Sarina and Heinke, Reinhard and Hosseini, Niyusha and Huyse, Mark and Köster, Ulli and Kudryavtsev, Yuri and Laatiaoui, Mustapha and Lica, Razvan and Magchiels, Goele and Manea, Vladimir and Merckling, Clement and Pereira, Lino M. C. and Raeder, Sebastian and Schumm, Thorsten and Sels, Simon and Thirolf, Peter G. and Tunhuma, Shandirai Malven and Van Den Bergh, Paul and Van Duppen, Piet and Vantomme, André and Verlinde, Matthias and Villarreal, Renan and Wahl, Ulrich},
	date = {2023-05},
    year = {2023},
	}

@article{kraemer_setup_2023,
title = {A setup for vacuum-ultraviolet spectroscopy of the $^{229}${T}h low-energy isomer},
journal = {Nuclear Instruments and Methods in Physics Research Section B: Beam Interactions with Materials and Atoms},
volume = {542},
pages = {1-3},
year = {2023},
issn = {0168-583X},
doi = {10.1016/j.nimb.2023.05.029},
url = {https://www.sciencedirect.com/science/article/pii/S0168583X23002288},
author = {Sandro Kraemer and Premaditya Chhetri and Silvia Bara and Arno Claessens and Hilde {De Witte} and Yens Elskens and Rafael Ferrer and Yuri Kudryavtsev and Simon Sels and Paul {Van Den Bergh} and Piet {Van Duppen}}
}

@phdthesis{kraemer_thesis_2022,
	author = {S. Kraemer and P. Van Duppen},
	title = {Vacuum-ultraviolet spectroscopy of the radiative decay of the low-energy isomer in $^{229}${T}h},
	school = {KU Leuven - Instituut voor Kern- en Stralingsfysica},
	year = {2022}
}

@article{zhang_frequency_2024,
	title = {Frequency ratio of the $^{229\mathrm{m}}${T}h nuclear isomeric transition and the $^{87}${S}r atomic clock},
	volume = {633},
	pages = {63--70},
	number = {8028},
	journal = {Nature},
	author = {Zhang, Chuankun and Ooi, Tian and Higgins, Jacob S. and Doyle, Jack F. and von der Wense, Lars and Beeks, Kjeld and Leitner, Adrian and Kazakov, Georgy A. and Li, Peng and Thirolf, Peter G. and Schumm, Thorsten and Ye, Jun},
	date = {2024-09},
    year = {2024}
}

@article{elwell_laser_2024,
	title = {Laser Excitation of the $^{229}${T}h Nuclear Isomeric Transition in a Solid-State Host},
	volume = {133},
	pages = {013201},
	number = {1},
	journal = {Physical Review Letters},
	shortjournal = {Phys. Rev. Lett.},
	author = {Elwell, R. and Schneider, Christian and Jeet, Justin and Terhune, J. E. S. and Morgan, H. W. T. and Alexandrova, A.N. and Tran Tan, H. B. and Derevianko, Andrei and Hudson, Eric R.},
	year = {2024}
}

@article{tiedau_laser_2024,
	title = {Laser Excitation of the {T}h-229 Nucleus},
	volume = {132},
	pages = {182501},
	number = {18},
	journal = {Physical Review Letters},
	shortjournal = {Phys. Rev. Lett.},
	author = {Tiedau, J. and Okhapkin, M. V. and Zhang, K. and Thielking, J. and Zitzer, G. and Peik, E. and Schaden, F. and Pronebner, T. and Morawetz, I. and De Col, L. Toscani and Schneider, F. and Leitner, A. and Pressler, M. and Kazakov, G. A. and Beeks, K. and Sikorsky, T. and Schumm, T.},
	date = {2024-04-29},
    year = {2024},
	}

@article{pineda_radiative_2025,
	title = {Radiative decay of the $^{229\mathrm{m}}${T}h nuclear clock isomer in different host materials},
	volume = {7},
	pages = {013052},
	number = {1},
	journal = {Physical Review Research},
	shortjournal = {Phys. Rev. Res.},
	author = {Pineda, S. V. and Chhetri, P. and Bara, S. and Elskens, Y. and Casci, S. and Alexandrova, A. N. and Au, M. and Athanasakis-Kaklamanakis, M. and Bartokos, M. and Beeks, K. and Bernerd, C. and Claessens, A. and Chrysalidis, K. and Cocolios, T. E. and Correia, J. G. and De Witte, H. and Elwell, R. and Ferrer, R. and Heinke, R. and Hudson, E. R. and Ivandikov, F. and Kudryavtsev, Yu. and Köster, U. and Kraemer, S. and Laatiaoui, M. and Lica, R. and Merckling, C. and Morawetz, I. and Morgan, H. W. T. and Moritz, D. and Pereira, L. M. C. and Raeder, S. and Rothe, S. and Schaden, F. and Scharl, K. and Schumm, T. and Stegemann, S. and Terhune, J. and Thirolf, P. G. and Tunhuma, S. M. and Van Den Bergh, P. and Van Duppen, P. and Vantomme, A. and Wahl, U. and Yue, Z.},
    year = {2025}
	}

@article{verlinde_alternative_2019,
  title={Alternative approach to populate and study the $^{229}${T}h nuclear clock isomer},
  author={Verlinde, M and Kraemer, S and Moens, J and Chrysalidis, K and Correia, JG and Cottenier, Stefaan and De Witte, H and Fedorov, DV and Fedosseev, VN and Ferrer, R and others},
  journal={Physical Review C},
  volume={100},
  number={2},
  pages={024315},
  year={2019},
  publisher={APS}
}

@article{thirolf_thorium_2024,
	title = {The thorium isomer $^{229\mathrm{m}}${T}h: review of status and perspectives after more than 50 years of research},
	volume = {233},
	pages = {1113--1131},
	number = {5},
	journal = {The European Physical Journal Special Topics},
	author = {Thirolf, Peter G. and Kraemer, Sandro and Moritz, Daniel and Scharl, Kevin},
	year = {2024}
}

@article{peik_nuclear_2003,
  title = {Nuclear laser spectroscopy of the 3.5 {eV} transition in $^{229}${T}h},
  author = {Peik, E. and Tamm, Chr},
  journal = {{EPL} (Europhysics Letters)},
  volume = {61},
  number = {2},
  pages = {181},
  year = {2003},
  month = {jan},
  issn = {0295-5075},
  doi = {10.1209/epl/i2003-00210-x}
}

@article{peik_nuclear_2021,
	title = {Nuclear clocks for testing fundamental physics},
	volume = {6},
	issn = {2058-9565},
	url = {https://dx.doi.org/10.1088/2058-9565/abe9c2},
	doi = {10.1088/2058-9565/abe9c2},
	
	pages = {034002},
	number = {3},
	journal = {Quantum Science and Technology},
	shortjournal = {Quantum Sci. Technol.},
	publisher = {{IOP} Publishing},
	author = {Peik, E. and Schumm, T. and Safronova, M. S. and Pálffy, A. and Weitenberg, J. and Thirolf, P. G.},
	urldate = {2024-12-16},
	
    journal = {Quantum Science and Technology},
    year = {2021},
    month = {apr},
	langid = {english}
}

@article{hiraki_Xray_2024,
  title={Controlling $^{229}${T}h isomeric state population in a {V}{U}{V} transparent crystal},
  author={Hiraki, Takahiro and Okai, Koichi and Bartokos, Michael and Beeks, Kjeld and Fujimoto, Hiroyuki and Fukunaga, Yuta and Haba, Hiromitsu and Kasamatsu, Yoshitaka and Kitao, Shinji and Leitner, Adrian and others},
  journal={Nature communications},
  volume={15},
  number={1},
  pages={5536},
  year={2024},
  publisher={Nature Publishing Group UK London}
}

@article{arakawa_arXiv_2026,
  title={Probing Ultralight Dark Matter at the Mega-Planck Scale with the Thorium Nuclear Clock},
  author={Arakawa, Jason and Doyle, Jack F and Fuchs, Elina and Higgins, Jacob S and Kirk, Fiona and Li, Kai and Ooi, Tian and Perez, Gilad and Ratzinger, Wolfram and Safronova, Marianna S and others},
  journal={arXiv preprint arXiv:2602.16804},
  year={2026}
}

@article{vonderwense_detection_2016,
  title={Direct detection of the $^{229}${T}h nuclear clock transition},
  author={von der Wense, Lars and Seiferle, Benedict and Laatiaoui, Mustapha and Neumayr, J{\"u}rgen B and Maier, Hans-J{\"o}rg and Wirth, Hans-Friedrich and Mokry, Christoph and Runke, J{\"o}rg and Eberhardt, Klaus and D{\"u}llmann, Christoph E and others},
  journal={Nature},
  volume={533},
  number={7601},
  pages={47--51},
  year={2016},
  publisher={Nature Publishing Group UK London}
}

@article{ooi_reproducibilty_2026,
  title={Frequency reproducibility of solid-state thorium-229 nuclear clocks},
  author={Ooi, Tian and Doyle, Jack F and Zhang, Chuankun and Higgins, Jacob S and Ye, Jun and Beeks, Kjeld and Sikorsky, Tomas and Schumm, Thorsten},
  journal={Nature},
  pages={1--7},
  year={2026},
  publisher={Nature Publishing Group UK London}
}

@article{guan_xray_2026,
  title = {X-Ray-Induced Quenching of the $^{229}\mathrm{Th}$ Clock Isomer in {C}a{F}$_{2}$},
  author = {Guan, Ming and Bartokos, Michael and Beeks, Kjeld and Fujimoto, Hiroyuki and Fukunaga, Yuta and Haba, Hiromitsu and Hiraki, Takahiro and Kasamatsu, Yoshitaka and Kitao, Shinji and Leitner, Adrian and Masuda, Takahiko and Nagasawa, Nobumoto and Okai, Koichi and Ogake, Ryoichiro and Pimon, Martin and Pressler, Martin and Sasao, Noboru and Schaden, Fabian and Schumm, Thorsten and Seto, Makoto and Shigekawa, Yudai and Shimizu, Kotaro and Sikorsky, Tomas and Tamasaku, Kenji and Takatori, Sayuri and Watanabe, Tsukasa and Yamaguchi, Atsushi and Yoda, Yoshitaka and Yoshimi, Akihiro and Yoshimura, Koji},
  journal = {Phys. Rev. Lett.},
  volume = {136},
  issue = {1},
  pages = {013203},
  numpages = {8},
  year = {2026},
  month = {1},
  publisher = {American Physical Society},
  doi = {10.1103/75bb-thn7},
  url = {https://link.aps.org/doi/10.1103/75bb-thn7}
}

@article{schaden_LIQ_2025,
  title = {Laser-induced quenching of the {T}h-229 nuclear clock isomer in calcium fluoride},
  author = {Schaden, F. and Riebner, T. and Morawetz, I. and De Col, L. Toscani and Kazakov, G. A. and Beeks, K. and Sikorsky, T. and Schumm, T. and Zhang, K. and Lal, V. and Zitzer, G. and Tiedau, J. and Okhapkin, M. V. and Peik, E.},
  journal = {Phys. Rev. Res.},
  volume = {7},
  issue = {2},
  pages = {L022036},
  numpages = {6},
  year = {2025},
  month = {5},
  publisher = {American Physical Society},
  doi = {10.1103/PhysRevResearch.7.L022036},
  url = {https://link.aps.org/doi/10.1103/PhysRevResearch.7.L022036}
}

@misc{hiraki2025lasermossbauer,
title={Laser {M}\"{o}ssbauer spectroscopy of $^{229}${T}h}, 
author={Takahiro Hiraki and Takahiko Masuda and Sayuri Takatori and Fabian Schaden and Michael Bartokos and Kjeld Beeks and Yuta Fukunaga and Andreas Grüneis and Ming Guan and Georgy Kazakov and Thomas LaGrange and Adrian Leitner and Ira Morawetz and Ryoichiro Ogake and Koichi Okai and Martin Pimon and Martin Pressler and Thomas Riebner and Noboru Sasao and Felix Schneider and Thorsten Schumm and Kotaro Shimizu and Luca Toscani de Col and Tomas Sikorsky and Akihiro Yoshimi and Koji Yoshimura},
      year={2025},
      eprint={2509.00041},
      archivePrefix={arXiv},
      primaryClass={nucl-ex},
      url={https://arxiv.org/abs/2509.00041}, 
}

@article{terhune2025photo,
  title = {Photoinduced quenching of the $^{229}\mathrm{Th}$ isomer in a solid-state host},
  author = {Terhune, J. E. S. and Elwell, R. and Tan, H. B. Tran and Perera, U. C. and Morgan, H. W. T. and Alexandrova, A. N. and Derevianko, Andrei and Hudson, Eric R.},
  journal = {Phys. Rev. Res.},
  volume = {7},
  issue = {2},
  pages = {L022062},
  numpages = {5},
  year = {2025},
  month = {6},
  publisher = {American Physical Society},
  doi = {10.1103/glzr-thyw},
  url = {https://link.aps.org/doi/10.1103/glzr-thyw}
}

@article{Catherall_ISOLDE_2017,
doi = {10.1088/1361-6471/aa7eba},
url = {https://doi.org/10.1088/1361-6471/aa7eba},
year = {2017},
month = {aug},
publisher = {IOP Publishing},
volume = {44},
number = {9},
pages = {094002},
author = {Catherall, R and Andreazza, W and Breitenfeldt, M and Dorsival, A and Focker, G J and Gharsa, T P and T J, Giles and Grenard, J-L and Locci, F and Martins, P and Marzari, S and Schipper, J and Shornikov, A and Stora, T},
title = {The ISOLDE facility},
journal = {Journal of Physics G: Nuclear and Particle Physics}
}

@article{li1980refractive,
  title={Refractive index of alkaline earth halides and its wavelength and temperature derivatives},
  author={Li, HH},
  journal={Journal of physical and chemical reference data},
  volume={9},
  number={1},
  pages={161--290},
  year={1980},
  publisher={American Institute of Physics for the National Institute of Standards and~…}
}

@article{hopkins2017abc,
  title={The ABC model of recombination reinterpreted: Impact on understanding carrier transport and efficiency droop in InGaN/GaN light emitting diodes},
  author={Hopkins, Margaret A and Allsopp, DWE and Kappers, MJ and Oliver, RA and Humphreys, CJ},
  journal={Journal of Applied Physics},
  volume={122},
  number={23},
  year={2017},
  publisher={AIP Publishing}
}

@article{morgan2024ICtheory,
  title = {Theory of Internal Conversion of the $^{229}\mathrm{Th}$ Nuclear Isomer in Solid-State Hosts},
  author = {Morgan, H. W. T. and Tran Tan, H. B. and Elwell, R. and Alexandrova, A. N. and Hudson, Eric R. and Derevianko, Andrei},
  journal = {Phys. Rev. Lett.},
  volume = {134},
  issue = {25},
  pages = {253801},
  numpages = {6},
  year = {2025},
  month = {Jun},
  publisher = {American Physical Society},
  doi = {10.1103/9s8f-hv1f},
  url = {https://link.aps.org/doi/10.1103/9s8f-hv1f}
}

@article{wang2012monte,
  title={Monte Carlo simulations of electron thermalization in alkali iodide and alkaline-earth fluoride scintillators},
  author={Wang, Zhiguo and Xie, YuLong and Campbell, Luke W and Gao, Fei and Kerisit, Sebastien},
  journal={Journal of Applied Physics},
  volume={112},
  number={1},
  year={2012},
  publisher={AIP Publishing}
}

@inproceedings{li2020study,
  title={Study of a large {C}a{F}$_2$ ({E}u) scintillating bolometer for neutrinoless double beta decay},
  author={Li, X and Kwon, DH and Tetsuno, K and Kim, I and Kim, HL and Lee, HJ and Yoshida, S and Kim, YH and Lee, MK and Umehara, S and others},
  booktitle={Journal of Physics: Conference Series},
  volume={1468},
  pages={012116},
  year={2020},
  organization={IOP Publishing}
}

@article{kawaguchi2018scintillation,
  title={Scintillation characteristics of Pr: Ca{F}$_2$ crystals for charged-particle detection},
  author={Kawaguchi, Noriaki and Kimura, Hiromi and Akatsuka, Masaki and Okada, Go and Kawano, Naoki and Fukuda, Kentaro and Yanagida, Takayuki},
  journal={Sens. Mater},
  volume={30},
  pages={1585--1590},
  year={2018}
}

@article{cang2020ionization,
  title={Ionization-density-dependent scintillation pulse shape and mechanism of luminescence quenching in LaBr3: Ce},
  author={Cang, Jirong and Fang, XinChao and Zeng, Zhi and Zeng, Ming and Liu, Yinong and Sun, Zhigang and Chen, Ziyun},
  journal={Physical Review Applied},
  volume={14},
  number={6},
  pages={064075},
  year={2020},
  publisher={APS}
}

@article{claessens2025_ionTh_PRA,
  title = {Thorium in hypersonic gas jets: Ionization potentials of {T}h and {T}h$^{+}$},
  author = {Claessens, A. and Ivandikov, F. and Brasseur, M. and Dragoun, A. and D\"ullmann, Ch. E. and Ferrer, R. and Kudryavtsev, Yu. and Palmeri, P. and Quinet, P. and Raeder, S. and Renisch, D. and Van den Bergh, P. and Van Duppen, P.},
  journal = {Phys. Rev. A},
  volume = {112},
  issue = {5},
  pages = {052810},
  numpages = {16},
  year = {2025},
  month = {Nov},
  publisher = {American Physical Society},
  doi = {10.1103/wqz7-ydrs},
  url = {https://link.aps.org/doi/10.1103/wqz7-ydrs}
}

@article{seiferle2017_lifetime_PRL,
  title = {Lifetime Measurement of the $^{229}\mathrm{Th}$ Nuclear Isomer},
  author = {Seiferle, Benedict and von der Wense, Lars and Thirolf, Peter G.},
  journal = {Phys. Rev. Lett.},
  volume = {118},
  issue = {4},
  pages = {042501},
  numpages = {5},
  year = {2017},
  month = {Jan},
  publisher = {American Physical Society},
  doi = {10.1103/PhysRevLett.118.042501},
  url = {https://link.aps.org/doi/10.1103/PhysRevLett.118.042501}
}

@article{Tkalya2015_radLifetime_PRC,
  title = {Radiative lifetime and energy of the low-energy isomeric level in $^{229}\mathrm{Th}$},
  author = {Tkalya, E. V. and Schneider, Christian and Jeet, Justin and Hudson, Eric R.},
  journal = {Phys. Rev. C},
  volume = {92},
  issue = {5},
  pages = {054324},
  numpages = {11},
  year = {2015},
  month = {Nov},
  publisher = {American Physical Society},
  doi = {10.1103/PhysRevC.92.054324},
  url = {https://link.aps.org/doi/10.1103/PhysRevC.92.054324}
}

@article{Kondev2026NDS,
  author  = {F. G. Kondev and J. K. Tuli and E. Browne},
  title   = {{E}valuated {N}uclear {S}tructure {D}ata {F}ile ({ENSDF})},
  journal = {Nuclear Data Sheets},
  volume  = {208},
  pages   = {397},
  year    = {2026}
}

@article{Morse2026NDS,
  author  = {C. Morse},
  title   = {{E}valuated {N}uclear {S}tructure {D}ata {F}ile ({ENSDF})},
  journal = {Nuclear Data Sheets},
  volume  = {204},
  pages   = {409},
  year    = {2026}
}

@article{Auranen2020NDS,
  author  = {K. Auranen and E. A. Mccutchan},
  title   = {{E}valuated {N}uclear {S}tructure {D}ata {F}ile ({ENSDF})},
  journal = {Nuclear Data Sheets},
  volume  = {168},
  pages   = {117},
  year    = {2020}
}

@article{Martin2007NDS,
  author  = {M. J. Martin},
  title   = {{E}valuated {N}uclear {S}tructure {D}ata {F}ile ({ENSDF})},
  journal = {Nuclear Data Sheets},
  volume  = {108},
  pages   = {1583},
  year    = {2007}
}

@mastersthesis{CasciThesis229Th,
  author      = {Simone Casci},
  title       = {Decay spectroscopy of $^{229}$Th implanted crystals for a nuclear clock},
  school      = {Università di {C}amerino},
  year        = {2024},
  type        = {Master's thesis},
  note        = {Supervisors: Alessandro Saltarelli; Co-supervisors: Piet Van Duppen, Silvia Bara, Yens Elskens}
}

@article{de2013influence,
  title={Influence of crystal mosaicity on axial channeling effects and lattice site determination of impurities},
  author={De Vries, Bart and Wahl, Ulrich and Ruffenach, Sandra and Briot, Olivier and Vantomme, Andr{\'e}},
  journal={Applied Physics Letters},
  volume={103},
  number={17},
  pages={172108},
  year={2013},
  publisher={AIP Publishing}
}

@article{caputo_2025_sensitivity,
  title = {Sensitivity of nuclear clocks to new physics},
  author = {Caputo, Andrea and Gazit, Doron and Hammer, Hans-Werner and Kopp, Joachim and Paz, Gil and Perez, Gilad and Springmann, Konstantin},
  journal = {Phys. Rev. C},
  volume = {112},
  issue = {3},
  pages = {L031302},
  numpages = {7},
  year = {2025},
  month = {Sep},
  publisher = {American Physical Society},
  doi = {10.1103/l29n-gt5j},
  url = {https://link.aps.org/doi/10.1103/l29n-gt5j}
}

@misc{toscani_2026_nuclear_clock_arxiv,
      title={A thorium-229 optical nuclear clock with feedback loop}, 
      author={L. Toscani De Col and T. Riebner and I. Morawetz and F. Schneider and N. Sempelmann and J. Schlachet-Lépinay and F. Schaden and M. Bartokos and G. A. Kazakov and K. Beeks and B. Gerstenecker and M. Pimon and S. Lahs and A. Hellerschmied and T. Lercher and J. Premper and A. Niessner and M. Matus and H. Denker and M. Cizek and O. Cip and V. Lal and G. Zitzer and V. Petrov and J. Tiedau and M. V. Okhapkin and E. Peik and T. Schumm},
      year={2026},
      eprint={2606.04997},
      archivePrefix={arXiv},
      primaryClass={physics.atom-ph},
      url={https://arxiv.org/abs/2606.04997}, 
}

\end{document}


\title{\textbf{Supplemental Material for: Exploring $\alpha$- and $\beta$-decay-induced quenching of the $^{229}$Th nuclear-clock isomer in solid-state hosts} 
}%
\title{\textbf{Exploring $\alpha$- and $\beta$-decay-induced quenching of the $^{229}$Th nuclear-clock isomer in solid-state hosts} 
}%

\author{Y.~Elskens}
\email{yens.elskens@kuleuven.be}
\affiliation{KU Leuven, Instituut voor Kern- en Stralingsfysica, 3001 Leuven, Belgium}

\author{M.~Athanasakis-Kaklamanakis}
\affiliation{KU Leuven, Instituut voor Kern- en Stralingsfysica, 3001 Leuven, Belgium}
\author{S.~Arasada Pradeep}
\affiliation{Institute of Physics -- Johannes Gutenberg University Mainz, 55099 Mainz, Germany}
\author{M.~Au}
\affiliation{CERN SY-STI, Geneva, 1211, Switzerland}
\author{S.~Bara}
\affiliation{KU Leuven, Instituut voor Kern- en Stralingsfysica, 3001 Leuven, Belgium}
\affiliation{Université de Caen Normandie, ENSICAEN, CNRS/IN2P3, LPC CAEN, F-14000 Caen, France}
\affiliation{Grand Accélérateur National d'Ions Lourds (GANIL), CEA/DRF-CNRS/IN2P3, F-14076 Caen, France}
\author{M.~Bartokos}
\affiliation{Vienna Center for Quantum Science and Technology, Atominstitut, TU Wien, 1020 Vienna, Austria}
\author{K.~Beeks}
\affiliation{Vienna Center for Quantum Science and Technology, Atominstitut, TU Wien, 1020 Vienna, Austria}
\author{C.~Bernerd}
\affiliation{KU Leuven, Instituut voor Kern- en Stralingsfysica, 3001 Leuven, Belgium}
\author{B.~Biesmans}
\affiliation{KU Leuven, Quantum Solid State Physics, 3001 Leuven, Belgium}
\author{S.~Casci}
\affiliation{KU Leuven, Instituut voor Kern- en Stralingsfysica, 3001 Leuven, Belgium}
\author{P.~Chhetri}%
\affiliation{Department of Chemistry -- TRIGA Site, Johannes Gutenberg-Universität Mainz, 55099 Mainz, Germany}
\author{K.~Chrysalidis}
\affiliation{CERN SY-STI, Geneva, 1211, Switzerland}
\author{A.~Claessens}
\affiliation{KU Leuven, Instituut voor Kern- en Stralingsfysica, 3001 Leuven, Belgium}
\author{T.~E.~Cocolios}
\affiliation{KU Leuven, Instituut voor Kern- en Stralingsfysica, 3001 Leuven, Belgium}
\author{J.~G.~Correia}
\affiliation{Centro de Ciências e Tecnologias Nucleares, Departamento de Engenharia e Ciências Nucleares, Instituto Superior Técnico, Universidade de Lisboa, 2695-066 Bobadela LRS, Portugal}

\author{A.~R.~G.~Costa}
\affiliation{Centro de Ciências e Tecnologias Nucleares, Departamento de Engenharia e Ciências Nucleares, Instituto Superior Técnico, Universidade de Lisboa, 2695-066 Bobadela LRS, Portugal}

\author{H.~De Witte}
\affiliation{KU Leuven, Instituut voor Kern- en Stralingsfysica, 3001 Leuven, Belgium}
\author{S.~B.~Diewald}
\affiliation{Department of Chemistry -- TRIGA Site, Johannes Gutenberg-Universität Mainz, 55099 Mainz, Germany}
\author{Ch.~E.~D\"ullmann}
\affiliation{Department of Chemistry -- TRIGA Site, Johannes Gutenberg-Universität Mainz, 55099 Mainz, Germany}
\affiliation{GSI Helmholtzzentrum für Schwerionenforschung GmbH, 64291 Darmstadt, Germany}
\affiliation{Helmholtz Institute Mainz, 55099 Mainz, Germany}

\author{R.~Ferrer}
\affiliation{KU Leuven, Instituut voor Kern- en Stralingsfysica, 3001 Leuven, Belgium}
\author{R.~Heinke}
\affiliation{CERN SY-STI, Geneva, 1211, Switzerland}
\author{G.~Holthoff}
\affiliation{Ludwig-Maximilians-Universit\"at M\"unchen, 85748 Garching, Germany}
\author{F.~Ivandikov}
\affiliation{KU Leuven, Instituut voor Kern- en Stralingsfysica, 3001 Leuven, Belgium}
\author{Yu.~Kudryavtsev}
\affiliation{KU Leuven, Instituut voor Kern- en Stralingsfysica, 3001 Leuven, Belgium}
\author{U.~K\"oster}
\affiliation{Institut Laue Langevin, 38042 Grenoble, France}
\author{S.~Kraemer}
\affiliation{KU Leuven, Instituut voor Kern- en Stralingsfysica, 3001 Leuven, Belgium}
\author{M.~Laatiaoui}
\affiliation{Department of Chemistry -- TRIGA Site, Johannes Gutenberg-Universität Mainz, 55099 Mainz, Germany}
\affiliation{Grand Accélérateur National d'Ions Lourds (GANIL), CEA/DRF-CNRS/IN2P3, F-14076 Caen, France}
\author{R.~Lica}
\affiliation{Horia Hulubei National Institute for R\&D in Physics and Nuclear Engineering, RO-077125 Bucharest, Romania}
\author{C.~Merckling}
\affiliation{IMEC, 3001 Leuven, Belgium}
\author{J.~Moens}
\affiliation{KU Leuven, Quantum Solid State Physics, 3001 Leuven, Belgium}
\author{I.~Morawetz}
\affiliation{Vienna Center for Quantum Science and Technology, Atominstitut, TU Wien, 1020 Vienna, Austria}
\author{D.~Moritz}
\affiliation{Ludwig-Maximilians-Universit\"at M\"unchen, 85748 Garching, Germany}
\author{L.~M.~C.~Pereira}
\affiliation{KU Leuven, Quantum Solid State Physics, 3001 Leuven, Belgium}
\author{S.~V.~Pineda}
\affiliation{KU Leuven, Instituut voor Kern- en Stralingsfysica, 3001 Leuven, Belgium}
\affiliation{Université de Caen Normandie, ENSICAEN, CNRS/IN2P3, LPC CAEN, F-14000 Caen, France}
\affiliation{Grand Accélérateur National d'Ions Lourds (GANIL), CEA/DRF-CNRS/IN2P3, F-14076 Caen, France}
\author{S.~Raeder}
\affiliation{GSI Helmholtzzentrum für Schwerionenforschung GmbH, 64291 Darmstadt, Germany}
\author{S.~Rothe}
\affiliation{CERN SY-STI, Geneva, 1211, Switzerland}
\author{S.~Sabrieva}
\affiliation{Vienna Center for Quantum Science and Technology, Atominstitut, TU Wien, 1020 Vienna, Austria}
\author{M.~Satrazani}
\affiliation{KU Leuven, Instituut voor Kern- en Stralingsfysica, 3001 Leuven, Belgium}
\author{F.~Schaden}
\affiliation{Vienna Center for Quantum Science and Technology, Atominstitut, TU Wien, 1020 Vienna, Austria}
\author{K.~Scharl}
\affiliation{Ludwig-Maximilians-Universit\"at M\"unchen, 85748 Garching, Germany}
\author{T.~Schumm}
\affiliation{Vienna Center for Quantum Science and Technology, Atominstitut, TU Wien, 1020 Vienna, Austria}
\author{S.~Stegemann}
\affiliation{CERN SY-STI, Geneva, 1211, Switzerland}
\author{J.~Stricker}
\affiliation{Department of Chemistry -- TRIGA Site, Johannes Gutenberg-Universität Mainz, 55099 Mainz, Germany}
\affiliation{Helmholtz Institute Mainz, 55099 Mainz, Germany}
\author{T.~Teschler}
\affiliation{Ludwig-Maximilians-Universit\"at M\"unchen, 85748 Garching, Germany}
\author{P.~G.~Thirolf}
\affiliation{Ludwig-Maximilians-Universit\"at M\"unchen, 85748 Garching, Germany}
\author{P.~Van den Bergh}
\affiliation{KU Leuven, Instituut voor Kern- en Stralingsfysica, 3001 Leuven, Belgium}
\author{P.~Van Duppen}
\affiliation{KU Leuven, Instituut voor Kern- en Stralingsfysica, 3001 Leuven, Belgium}
\author{A.~Vantomme}
\affiliation{KU Leuven, Quantum Solid State Physics, 3001 Leuven, Belgium}
\author{R.~Villarreal}
\affiliation{KU Leuven, Quantum Solid State Physics, 3001 Leuven, Belgium}
\author{L. von der Wense}
\affiliation{Institute of Physics -- Johannes Gutenberg University Mainz, 55099 Mainz, Germany}
\author{U.~Wahl}
\affiliation{Centro de Ciências e Tecnologias Nucleares, Departamento de Engenharia e Ciências Nucleares, Instituto Superior Técnico, Universidade de Lisboa, 2695-066 Bobadela LRS, Portugal}
\author{Y.~Wang}
\affiliation{Institute of Physics -- Johannes Gutenberg University Mainz, 55099 Mainz, Germany}
\author{M.~Wiesinger}
\affiliation{Ludwig-Maximilians-Universit\"at M\"unchen, 85748 Garching, Germany}
\author{Z.~Yue}
\affiliation{CERN SY-STI, Geneva, 1211, Switzerland}
\author{F.~Zacherl}
\affiliation{Institute of Physics -- Johannes Gutenberg University Mainz, 55099 Mainz, Germany}
\date{\today}
\maketitle
\subsection{Bateman equations modified for quenching}
\subsubsection*{$\beta$-decay-induced quenching}
The time behavior of the radioactive ions from the $^{229}$Fr$\rightarrow^{229}$Ra$\rightarrow^{229}$Ac decay chain is described by the Bateman equations
\begin{subequations}
    \begin{align}
    \dv{N_\mathrm{Fr}}{t} &= R_\mathrm{Fr}-\lambda_\mathrm{Fr}N_\mathrm{Fr}\,,\label{eq: bateman_Fr}\\
    \dv{N_\mathrm{Ra}}{t} &= R_\mathrm{Ra}+\lambda_\mathrm{Fr}N_\mathrm{Fr} - \lambda_{\mathrm{Ra}}N_\mathrm{Ra}\,, \label{eq: bateman_Ra}\\
    \dv{N_\mathrm{Ac}}{t} &= R_\mathrm{Ac} + \lambda_\mathrm{Ra}N_\mathrm{Ra} - \lambda_\mathrm{Ac}N_\mathrm{Ac}\,, \label{eq: bateman_Ac}
\end{align}
\end{subequations}

where $R_\mathrm{Fr}$, $R_\mathrm{Ra}$ and $R_\mathrm{Ac}$ represent the implantation rates for $^{229}$Fr, $^{229}$Ra and $^{229}$Ac, respectively. These implantation rates were inferred from $\gamma$-ray spectroscopy experiments, from which it was found that the implantation rates for $^{229}$Fr and $^{229}$Ac were two orders of magnitude lower compared to $^{229}$Ra. As a result, the feeding term of the $^{229}$Ac equation was solely attributed to the decay of $^{229}$Ra, and the contributions of $^{229}$Fr to the decay chain  were considered negligible altogether.

The isomer is populated via the $\beta$-decay of $^{229}$Ac, with a branching ratio (BR) in the range of $14\%$ to $93\%$\cite{verlinde_alternative_2019}. Only an a priori unknown fraction (the so-called radiative-decay fraction, RDF) of the isomers will decay radiatively, and therefore have a chance of being detected by the VUV spectrometer. 

In the present phenomenological model, RDF should not be interpreted as the fraction of nuclei residing in defect configurations for which the ideal host band gap is strictly preserved. Rather, RDF denotes the fraction of $^{229\mathrm{m}}$Th nuclei whose static local electronic configuration remains radiatively active, i.e. for which the local electronic, defect, or excitonic spectrum does not provide an efficient non-radiative decay channel at the nuclear transition energy $E_{\mathrm{iso}}$. Static configurations that quench the isomer promptly, or on a time scale not resolved in the present experiment, are absorbed into RDF and therefore only affect the overall normalization of the observed signal.

The Bateman equations describing the time behavior of this radiatively active ensemble, including additional activity-dependent quenching, then become 
\begin{subequations}
    \begin{align}
    \dv{N^{(1)}_\mathrm{iso}}{t} &= \xi\times\mathrm{RDF}\times \mathrm{BR}\times\lambda_\mathrm{Ac}N_\mathrm{Ac}\notag\\ &\quad- [\lambda_0+\lambda^{(1)}_Q(t)]N^{(1)}_\mathrm{iso}\,,\label{eq: N1}\\
    \dv{N^{(2)}_\mathrm{iso}}{t} & = (1-\xi) \times\mathrm{RDF}\times \mathrm{BR}\times\lambda_\mathrm{Ac}N_\mathrm{Ac}\notag\\ &\quad - [\lambda_0+\lambda^{(2)}_Q(t)]N^{(2)}_\mathrm{iso}\,.\label{eq: N2}
\end{align}
\end{subequations}
The data allowed to distinguish the time behavior of up to two different fractions of isomers: a fraction $\xi$ whose sensitivity to quenching is encapsulated in the activity-dependent decay constant $\lambda^{(1)}_Q(t)$, and the remaining fraction $(1-\xi)$ which undergoes quenching with $\lambda^{(2)}_Q(t)$. For both decay constants a linear relationship with the total $\beta$ activity is assumed, with characteristic proportionality factors $q^{(1)}_\beta$ and $q^{(2)}_\beta$, respectively.
\begin{subequations}
    \begin{align}
    \lambda^{(1)}_Q(t)&= q^{(1)}_\beta\times(\lambda_\mathrm{Ra}N_\mathrm{Ra}+\lambda_\mathrm{Ac}N_\mathrm{Ac})\,,\\
    \lambda^{(2)}_Q(t)&= q^{(2)}_\beta\times(\lambda_\mathrm{Ra}N_\mathrm{Ra}+\lambda_\mathrm{Ac}N_\mathrm{Ac})\,,
\end{align}
\end{subequations}
Here the expressions for $N_\mathrm{Ra}$ and $N_\mathrm{Ac}$ were determined from the respective Bateman equations~\cref{eq: bateman_Ra,eq: bateman_Ac}. To describe the total observed radiative-decay rate ($r_\mathrm{obs}$), \cref{eq: N1,eq: N2} are combined as follows
\begin{equation}
    r_\mathrm{obs} = \varepsilon_I\times \lambda_0(N^{(1)}_\mathrm{iso}+N^{(2)}_\mathrm{iso})\,,
\end{equation}
where $\varepsilon_I$ represents the total instrumental efficiency, taking into account the collection efficiency of the mirrors, the diffraction grating, entrance- and exit slits, and the PMT. 

Since $\varepsilon_I$, BR and RDF are all a priori unknown and only result in an overall scaling of the observed radiative-decay rate, these parameters were combined in a single overall-efficiency parameter $\eta=\varepsilon_I\times\mathrm{RDF}\times\mathrm{BR}$. This reduces the analysis to inferring the posterior distributions of four parameters: the two quenching probabilities  $q^{(1)}_\beta$ and $q^{(2)}_\beta$, the fraction $\xi$ of isomers with quenching probability $q^{(1)}_\beta$, and the overall efficiency $\eta$.

However, the sensitivity in the CaF$_2$ crystals to quenching was too low for the data to constrain a model where both quenching parameters are nonzero. As a result $q^{(2)}_\beta$ was fixed to zero. The sensitivity in the CaF$_2$ UV-grade crystal was so low, that only a single-fraction model could be constrained, hence also fixing $\xi=1$.  
\subsubsection*{$\alpha$-decay-induced quenching}
When studying the effects of $\alpha$ decay on the quenching of the radiative decay, the total $\alpha$ activity within the $^{220}$Fr decay chain and the total $\beta$ activity of both the $A=229$ and the $A=220$ chain need to be taken into account.

\begin{subequations}
    \begin{align}
        \dv{N_\mathrm{Fr}}{t} &= R_\mathrm{Fr} - \lambda_\mathrm{Fr}N_\mathrm{Fr}\,,\\
        \dv{N_\mathrm{At}}{t} & = \lambda_\mathrm{Fr}N_\mathrm{Fr} - \lambda_\mathrm{At}N_\mathrm{At}\,,\\
        \dv{N_\mathrm{Bi}}{t} & = \lambda_\mathrm{At}N_\mathrm{At} - \lambda_\mathrm{Bi}N_\mathrm{Bi}\,,\\
        \dv{N_\mathrm{Po}}{t} &= 0.64\,\lambda_\mathrm{Bi}N_\mathrm{Bi} - \lambda_\mathrm{Po}N_\mathrm{Po}\,,\\
        \dv{N_\mathrm{Tl}}{t} & = 0.36\, \lambda_\mathrm{Bi}N_\mathrm{Bi} - \lambda_\mathrm{Tl}N_\mathrm{Tl}\,. 
    \end{align}
\end{subequations}
The activity-dependent decay constants then become
\begin{subequations}
    \begin{align}
    \lambda^{(1)}_Q(t)&= q^{(1)}_\alpha\,A_\alpha + q^{(1)}_\beta\,A_\beta \,,\\
    \lambda^{(2)}_Q(t)&= q^{(2)}_\alpha\,A_\alpha + q^{(2)}_\beta\,A_\beta\,,
\end{align}
\end{subequations}
where 
\begin{subequations}
    \begin{align}
    A_\alpha &= \lambda_\mathrm{Fr}N_\mathrm{Fr}+\lambda_\mathrm{At}N_\mathrm{At} +0.36\,\lambda_\mathrm{Bi}N_\mathrm{Bi} + \lambda_\mathrm{Po}N_\mathrm{Po}\,,\\
    A_\beta &= 0.64\, \lambda_\mathrm{Bi}N_\mathrm{Bi} + \lambda_\mathrm{Tl}N_\mathrm{Tl} + \lambda_\mathrm{Ra}N_\mathrm{Ra} + \lambda_\mathrm{Ac}N_\mathrm{Ac}\,.
\end{align}
\end{subequations}

\subsection{Emission channeling in M{g}F$_2$}
Similar to the CaF$_2$ data presented in the main text of the paper, an emission channeling (EC) experiment with MgF$_2$ was carried out, but instead of $^{231}$Th with $^{229}$Ac, the precursor isotope of $^{229}$Th. Similar experiments have previously also been performed in CaF$_2$, as reported in Ref. ~\cite{verlinde_alternative_2019}. The similarity between the EC patterns of $^{229}$Ac in MgF$_2$ and those previously observed for $^{229}$Ac in CaF$_2$ suggests that $^{229}$Ac occupies an analogous cation-substitutional lattice site in MgF$_2$, corresponding to the Mg site. Although this qualitative comparison does not constitute a direct lattice-site determination, it provides a strong indication of the lattice environment expected for the daughter isotope $^{229}$Th following the decay of $^{229}$Ac in MgF$_2$. The crystallographic structure of MgF$_2$ is similar to the one of CaF$_2$, although it is tetragonal (Rutile structure) and not cubic: while both CaF$_2$ and MgF$_2$ have orthogonal main crystallographic axes, CaF$_2$ has a single, cubic lattice constant of $a=5.463\,\si{\angstrom}$, while MgF$_2$ possesses different lattice constants of $a=b=4.621\,\si{\angstrom}$, and $c=3.053\,\si{\angstrom}$. The fact that  MgF$_2$ is tetragonal, made it impossible to use the current implementation of the ``manybeam" code to simulate the theoretical EC patterns expected from this structure. 
Hence, in the following we present only a qualitative analysis of the obtained EC data in MgF$_2$. 

\begin{figure}[t]
    \centering

    \includegraphics[width=0.8\linewidth]{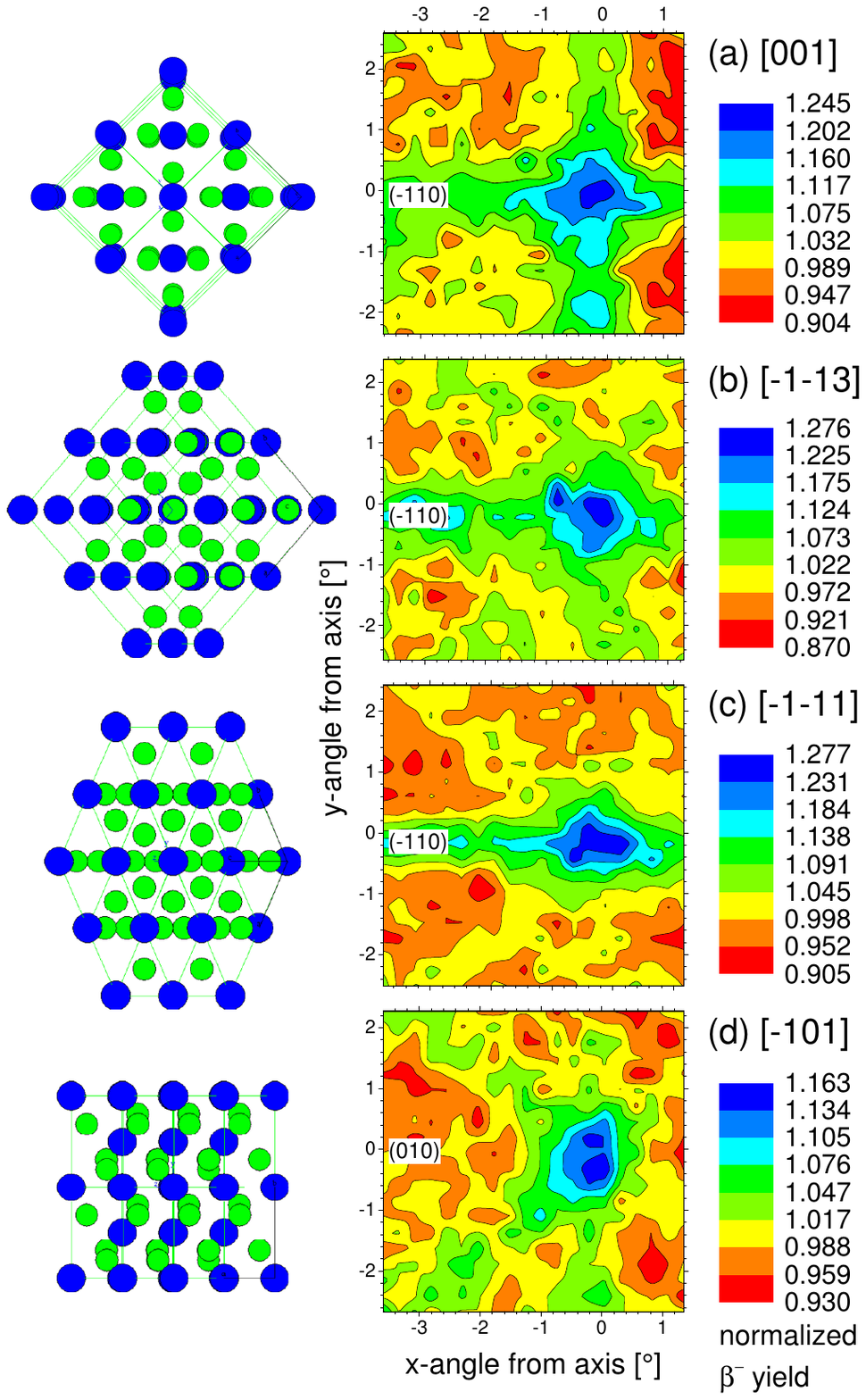}

    \caption{Right: Experimental emission channeling patterns of $\beta^{-}$ particles emitted by $^{229}$Ac in the MgF$_2$ single crystal, measured in the as-implanted state along the indicated crystallographic directions [001], [-1-13], [-1-11], and [-101]. Left: Projections of the MgF$_2$ lattice along these crystallographic directions: Mg atoms are shown in blue while F atoms are in light green.}
    \label{fig:EC_MgFAc01}
\end{figure}

Supplementary Fig.~\ref{fig:EC_MgFAc01} shows the angular-dependent experimental $\beta^{-}$ emission patterns resulting from $^{229}$Ac in MgF$_2$, in the vicinity of its main four crystallographic axes [001], [-1-13], [-1-11], and [-101]. The fact that channeling effects are observed along all of these directions, indicates that a significant fraction of $^{229}$Ac occupies substitutional lattice sites. It is reasonable to expect that these are substitutional Mg sites, similar to $^{229}$Ac on substitutional Ca sites in CaF$_2$ ~\cite{verlinde_alternative_2019}. 

\bibliographystyle{apsrev4-2}
\bibliography{supplemental}